\documentclass[12pt]{iopart}
\pdfoutput=1
\usepackage{amsfonts}
\usepackage{graphicx}
\usepackage{iopams}
\usepackage[mathscr]{euscript}

\usepackage{xcolor}
\usepackage{bm}
\usepackage{subfigure}
\usepackage{textcomp}
\usepackage{cite}
\usepackage{float}
\usepackage{soul}
\usepackage{stackrel}
\usepackage{ulem}
\usepackage{multirow}

\expandafter\let\csname equation*\endcsname\relax
\expandafter\let\csname endequation*\endcsname\relax
\usepackage{amsmath,amssymb,mathtools}

\allowdisplaybreaks[1]

\makeatletter
\newenvironment{sqcases}{%
  \matrix@check\sqcases\env@sqcases
}{%
  \endarray\right.%
}
\def\env@sqcases{%
  \let\@ifnextchar\new@ifnextchar
  \left\lbrack
  \def\arraystretch{1.2}%
  \array{@{}l@{\quad}l@{}}%
}

\usepackage{stackengine}

\usepackage{booktabs}

\usepackage{tikz}
\usepackage{esint}
\usetikzlibrary{arrows,arrows.meta}
\newcommand\semiInt[1][1]{%
    \begin{tikzpicture}[scale=0.2]
        \coordinate (center) at (0.2,0.55);
        \draw[black,-{>[scale=0.6]}] (0, 0.0) + (center) arc (180:0:0.5);
       $\int$
    \end{tikzpicture}
    }

\expandafter\let\csname equation*\endcsname\relax
\expandafter\let\csname endequation*\endcsname\relax
\usepackage{amsmath,amssymb}

\usepackage[colorlinks,linkcolor=blue,urlcolor=blue,citecolor=blue]{hyperref}
\usepackage{soul,xcolor}
\allowdisplaybreaks

\pdfoutput=1
\usepackage{esint}

\usepackage{hyperref}
\usepackage{color}
\definecolor{dvo}{rgb}{0.7,0.2,0.2}
\definecolor{dgreen}{rgb}{0,0.7,0}

\def\be{\begin{equation}}
\def\ee{\end{equation}}
\def\bea{\begin{eqnarray}}
\def\eea{\end{eqnarray}}

\begin{document}

\title{Resetting with stochastic return through linear confining potential}

\author{Deepak Gupta$^{1}$, Arnab Pal$^{2}$, Anupam Kundu$^{3}$ \footnote{anupam.kundu@icts.res.in}}

\address{\noindent {$^{1}$Dipartimento di Fisica `G. Galilei', INFN, Universit\`a di Padova, Via Marzolo 8, 35131 Padova, Italy}}

\address{\noindent {$^{2}$School of Chemistry, The Center for Physics and Chemistry of Living Systems, Tel Aviv University, Tel Aviv 6997801, Israel}}

\address{\noindent {$^{3}$International Centre for Theoretical Sciences, Tata Institute of Fundamental Research, Bengaluru 560089, India}}

\date{\today}

\begin{abstract}
{We consider motion of an overdamped Brownian particle subject to stochastic resetting in one dimension. In contrast to the usual setting where the particle is instantaneously reset to a preferred location (say, the origin), here we consider a finite time resetting process facilitated by an external linear potential $V(x)=\lambda|x|~ (\lambda>0)$. When resetting occurs, the trap is switched on and the particle experiences a force $-\partial_x V(x)$ which helps the particle to return to the resetting location. The trap is switched off as soon as the particle makes a first passage to the origin. Subsequently, the particle resumes its free diffusion motion and the process keeps repeating. In this set-up, the system attains a non-equilibrium steady state. We study the relaxation to this steady state by analytically computing the position distribution of the particle at all time and then analysing this distribution using the spectral properties of the corresponding Fokker-Planck operator.  As seen for the instantaneous resetting problem, we observe a `cone spreading' relaxation with travelling fronts such that there is an inner core region around
the resetting point that reaches the steady state, while the region outside the core still grows ballistically with time. {In addition to the unusual relaxation phenomena, 
we compute the large deviation functions associated to the corresponding probability density and find that the large deviation functions describe a dynamical transition similar 
to what is seen previously in case of instantaneous resetting.} Notably, our method, based on spectral properties, complements the existing renewal 
formalism and reveals the intricate mathematical structure responsible for such relaxation phenomena. We verify our analytical results against extensive numerical simulations.}
\end{abstract}

\noindent{\bf Keywords:} {Stochastic resetting, Brownian motion, Fokker-Planck equation}

\maketitle

\section{Introduction}
\label{intro}
Stochastic resetting is a mechanism that interrupts an ongoing process at random times and resets it to some pre-determined state only to recommence all over again \cite{Review}. Such tendencies are only natural and appear in many search processes e.g., animal foraging \cite{Restart-Search1,HRS}, randomized computer search algorithms \cite{algorithm,Luby}, facilitated diffusion \cite{FTD}, and single molecular search processes such as backtracking by RNA polymerases \cite{bio},  motor driven intracellular transport \cite{Restart-Search2},  cytoneme based search of morphogen \cite{Restart-Search3} and enzymatic reactions \cite{ReuveniEnzyme1,ReuveniEnzyme3}. Beyond search strategies, the notion of resetting has also been found ubiquitously in several other contexts e.g., in population dynamics \cite{population-1}, queuing system \cite{queue-1}, extreme and catastrophic events \cite{Montanari,extreme,cat-ev}.

Systematic theoretical study of resetting in the physics literature started with the canonical Evans-Majumdar model for diffusion with stochastic resetting \cite{Restart1} and since then the subject has been a focal point of recent studies. Simply put, resetting brings a particle back to a preferred location persistently thus putting it into an effective confinement. This generates a non-zero probability current in the system and eventually a 
non-equilibrium steady state is attained \cite{Restart2,Pal-potential,Kirone,transport1,transport2,Pal-time-dep,Durang,path}. Steady state properties under resetting have also been studied in a wide spectrum of stochastic processes namely anomalous diffusion \cite{anamolous-1,anamolous-3}, underdamped process \cite{underdamped}, random acceleration process \cite{RAP}, scaled Brownian motion \cite{scaled}, continuous time random walk \cite{CTRW1,CTRW2}, L\'evy flights \cite{Levy-flight}, telegraphic process \cite{telegraphic}, run and tumble motion \cite{RT,RTP} and others \cite{local-time,restart_conc17,magnetic,SRRW,comb}.
Steady state properties under resetting were also studied in many-body interacting systems such as fluctuating interface \cite{KPZ}, {spin system \cite{Ising}} and exclusion process \cite{SEP,ASEP}.

Another hallmark property of resetting is to expedite the completion of a first passage process which otherwise would hinder e.g., resetting renders the mean first passage time finite for a diffusive search process \cite{Restart1,Restart2,Pal-time-dep}. Regularization by resetting happens by taking advantage of the large stochastic fluctuations in the underlying first passage processes and this property was extensively studied in diffusive systems \cite{bounded,VV,ORR,Landau,Peclet,log,Gupta,cost,cost2}, generic stochastic processes \cite{Chechkin,ReuveniPRL,PalReuveniPRL,branching,Belan} and nonlinear dynamical systems such as chaotic Lorenz system \cite{nonlinear}. Resetting has also been studied in the context of stochastic thermodynamics \cite{thermo1,thermo2,thermo3,thermo4,thermo5,sp-lm}, large deviation \cite{restart_conc17, Hollander_2019} and in quantum systems \cite{quantum1}.

Majority of the studies to date in the field assumed the resetting process to take zero time \cite{Review}. However, in practice, getting from one place to another for the particle always takes non-zero time.
This idea has recently been augmented with a series of theoretical studies which attempted to account for such realistic situations. To be more specific, in the set-up of diffusion with resetting, the particle is reset/returned to the origin at a finite time making use of various deterministic protocols e.g., returns with a constant or space dependent velocity \cite{HRS,invariance1,invariance2,return4,return3,return3-1}, constant or space dependent acceleration \cite{return3,return3-1}, and returns with finite time \cite{refractory}. This intuition also stems from the first experimental case study of the resetting phenomena \cite{expt-1} (also see \cite{expt-2} which followed after) where it was shown why it is only natural to consider resetting processes with non-instantaneous returns.

Our motivation to the current paper complements the idea of deterministic return protocols with the stochastic ones since in reality, a deterministic driving is never perfect and will always be accompanied by uncontrollable random fluctuations. Taking  these facts into account, we recently have proposed the idea of stochastic resetting followed by stochastic return in \cite{st-rt} {(also see another recent study \cite{irp} that proposes similar ideas of diffusion
with stochastic returns using intermittent resetting potentials)}. To illustrate the set-up, let us again consider
the Brownian particle which diffuses in one dimension until a random resetting time. At that very moment, an external potential centered at the resetting location, is turned on and the particle starts experiencing a force towards the potential minimum (i.e., the resetting location). Once the particle hits the resetting location for \textit{the first time}, the trap is switched off immediately, and the particle resumes its diffusing motion. Thus, motion of the particle can viewed as a compound process which is a superposition of the exploration and return phase. In Ref.~\cite{st-rt}, we focused on the steady state and discussed shapes and forms of the position density in each phase analytically for different choices of external confining potentials. In this paper, we go beyond the steady state analysis to study the time dependent problem in whole generality. We provide a comprehensive method to compute the exact time dependent density, and then apply the framework to  a set-up where a Brownian particle is returned to the origin using a linear trap potential i.e., $V(x)=\lambda|x|$ ( $\lambda>0$). We first provide the exact form of the position density and then study its large time relaxation behaviour.

Often relaxation of a system to a stationary state is described by the exponential relaxation through spectral representation of the Fokker-Planck operator where the relaxation occurs homogeneously over the full available space with uniform rate. In contrast, for the canonical diffusion with instantaneous resetting model, it was demonstrated using the renewal structure that the relaxation to the steady state occurs in-homogeneously \cite{transport1}.
More elaborately, it was shown that as time progresses an inner core region around the resetting point reaches the steady state, while the region outside the core is still transient. The boundaries of the core region are seen to grow with time  linearly \cite{transport1}. In this paper we show that a similar `cone spreading' type relaxation with travelling fronts also occurs in the case of resetting with stochastic return using a linear trap. We find that details of the core e.g., the speed of its boundaries depends on the relative strength between the resetting rate $r$ and the potential $\lambda$. We solve the Fokker-Planck equation using Laplace transform method which reveals the spectral properties of the operator. Analysing the spectral properties through inverse Laplace transform we study the relaxation to the steady state at large time. This approach does not only complement the existing renewal formalism but also proves to be effective where a renewal framework is less tractable. The current problem at our hand exactly depicts such a scenario.

{Finally, Finally, we analyse the large deviation property of the particle's position density. 
For the instantaneous case, the large deviation function (LDF) was studied in \cite{transport1} where it has been described to demonstrate a dynamical transition in the density with time when looked at a fixed position.  
{In the present case of non-instantaneous resetting, we also find a similar dynamical transition} described by appropriate LDFs. In particular, we find that the position 
density satisfies a large deviation form with an explicit LDF. The form of the LDF depends on the relative strength between the resetting rate $r$ and the potential strength $\lambda$. 
In each case, the LDF can be interpreted to describe a dynamical transition occurring at a fixed position similar to what one observes in the instantaneous case \cite{transport1}.}


The structure of the paper is organized as follows. In Sec. \ref{model}, we briefly introduce the model and discuss the governing Fokker-Planck equations for the process. In Sec. \ref{sol-lap}, we sketch out the method to solve these equations in Laplace space. Sec. \ref{pb-t} contains the exact results for the probability density function of the particle at all times. Next in Sec. \ref{PD-PR}, we compute the probability for the system to be in the two phases. Sec. \ref{relaxation} is dedicated to the detailed computation of the relaxation of the total density to its steady state. We summarise our paper in Sec. \ref{summ}. Some details of the calculations are reserved for the Appendix.

\section{Model}
\label{model}
We consider motion of a Brownian particle interrupted by stochastic resetting in one-dimension. We refer the uninterrupted motion of the particle as the \textit{exploration or diffusive phase}. {Resetting} occurs at a rate $r$  {and at each resetting event,}
we switch on a  trap that creates a potential $V(x)$ centered (i.e., has its global minimum) at the resetting location $x_r$.  Thus, {after every resetting event} the particle experiences a force $F(x) = -\partial_x V(x)$  which facilitates the particle to return to the resetting location. This motion will be referred as the \textit{return phase}. The return phase persists until the particle hits the resetting location for the first time and at that moment the trap is switched off.
This marks the end of the return phase, and the particle resumes its motion in the (diffusive) exploration phase. In what follows, we will set the resetting location as the origin $x_r=0$ without loss of any generality. 

{Let $x(t)$ be the position of the particle at time~$t$.  The  evolution of the particle in the subsequent time interval $dt$ depends on which phase the particle is in. If the particle is in the exploration phase then within time interval $dt$, a resetting event can occur with probability $rdt$ and if so happens the external potential is switched on, and the particle starts moving in the force field. Otherwise, the particle continues to perform the free diffusion with probability $(1-rdt)$. On the other hand if at time $t$ the particle is in the return phase then it moves in force field until it reaches the origin where the free diffusion starts again. The evolution can be written as  following:
 \begin{align}
 \begin{split}
& \text{In~exploration~phase}  \\
&~~~~~~~~~~~~
x(t+dt) = 
\begin{sqcases}
x(t)+\sqrt{2D}~dW(t),~~&\text{with~prob.}~(1-rdt)\\
x(t)+F(x(t))dt + \sqrt{2D}~dW(t),~~&\text{with~prob.}~rdt
\end{sqcases}\\
&\text{In~return~phase}  \\
&~~~~~~~~~~~~
x(t+dt) = x(t)+F(x(t))dt + \sqrt{2D}~dW(t),
\end{split}
\label{eom}
 \end{align}
 where $dW(t)=\int_t^{t+dt}dt'\eta(t')$ with $\eta(t)$ being a zero-mean Gaussian white noise and $D$ is the diffusion constant.}

In each realization, the particle starts from the origin in the exploration phase, and has multiple switching between the two phases along a long trajectory of  {duration} $t$. 
Thus at any observation time, the particle can be  {found} at random in {either of the two phases.}   
Let  $\rho_D(x,t)$ and $\rho_R(x,t)$  {denote} the probability density {functions} of finding the particle at position $x$ at time $t$ in the exploration and return phase respectively. 
Thus, the probability density function for the compound process at time $t$ is given by \cite{st-rt}
\bea
\rho(x,t)=\rho_D(x,t)+\rho_R(x,t),
\eea
which is normalized $\int^{+\infty}_{-\infty}~dx~ \rho(x,t)=1$. On the other hand, the individual phase densities are normalized to the time dependent probabilities to be in each of the phases. To be specific, $p_D(t)=\int^{+\infty}_{-\infty}~dx~ \rho_D(x,t)$ and $p_R(t)=\int^{+\infty}_{-\infty}~dx~ \rho_R(x,t)$ are the time dependent probabilities to find the particle in the exploration and return phases respectively, with the normalization $p_D(t)+p_R(t)=1$.

In a recent work \cite{st-rt},  {a general framework 
has been developed} {to} describe   stochastic processes under stochastic returns. Following this formalism, and translating to our set-up, we get the following coupled Fokker-Planck (FP) equations for the densities
\begin{align}
\partial_t \rho_D(x,t)=-\partial_x J_D(x,t)-r \rho_D(x,t)+\delta(x) \mathscr{J}_R(0,t),\label{rhoM-eqns}  \\
\partial_t \rho_R(x,t)=-\partial_x J_R(x,t)+r \rho_D(x,t)-\delta(x) \mathscr{J}_R(0,t), \label{rhoR-eqns} 
\end{align}
where $J_D(x,t)=-D \partial_x \rho_D(x,t)$ and $J_R(x,t)= F(x) \rho_R(x,t)-D \partial_x \rho_R(x,t)$
are the fluxes in {the  {exploration} and return} phases respectively. {Note here},
$\mathscr{J}_R(0,t)=J_R(0^-,t)-J_R(0^+,t)$
is the net flux of returning particles those enter at the origin from the negative and positive side, and    
{immediately switch their motion}  {to the exploration} phase. Since all the particles  {initially} start {their motion} in {exploration}   phase {(which is free diffusion),} we have
\bea
\rho_D(x,0)=\delta(x)~,~~\rho_R(x,0)=0.
\label{initial}
\eea
{Interestingly, the return phase density can be written as a sum of the densities $\rho_R^{+}(x,t)$ and $\rho_R^{-}(x,t)$ which describe the particles returning from $x>0$ and $x<0$, respectively, such that \cite{st-rt}}
\bea
   \rho_R(x,t)=\rho_R^-(x,t) \Theta(-x)+\rho_R^+(x,t) \Theta(x),
   \label{superposition}
\eea
where $\Theta(x)$ represents the Heaviside step function. This decomposition is only natural since the returning particles can either be in the positive or negative $x$ and switching between them is {possible only via the exploration} phase. Thus, the origin acts as an effective absorbing boundary for the return {probability densities} such that
\bea
\rho_R^{\pm}(0,t)=0.
\label{rhoR-abs}
\eea
Furthermore, since the probability can not get accumulated at infinity at any time for motions in both phases, we have the natural boundary conditions 
\begin{subequations}
\begin{align}
\lim_{x \to  \pm \infty} \rho_D(x,t) =0,\label{rhoD-infinity} \\
\quad \lim_{x \to  \pm \infty} \rho_R^{\pm}(x,t)=0.
\label{rhoR-infinity}
\end{align}
\end{subequations}
Finally, the {exploration} phase must be continuous at the origin implying
\begin{align}
\rho_D(0^{+},t)=\rho_D(0^{-},t).
\label{matching}
\end{align} 
We now {proceed} to solve the set of time dependent equations (namely Eqs. \eqref{rhoM-eqns} and \eqref{rhoR-eqns}) {along with the initial conditions in Eq.~(\ref{initial}) and boundary conditions in Eqs.~(\ref{rhoR-abs})-(\ref{matching}) discussed above}.

\section{Solution of the master equations}
\label{sol-lap}
In the {previous} section, we have presented the FP Eqs. \eqref{rhoM-eqns} and \eqref{rhoR-eqns} that govern the motion of particles in the  {exploration} and the return phase. To solve them, we first perform the following Laplace transforms 
$\tilde \rho_{D,R}(x,s)=\int_0^\infty~dt~e^{-s t}~\rho_{D,R}(x,t)$ and $\tilde j(s)=\int_0^\infty~dt~e^{-s t}~ \mathscr{J}_R(0,t)$ on both sides of Eqs.~\eqref{rhoM-eqns} and \eqref{rhoR-eqns} and get 
\begin{align}
D\dfrac{\partial^2 \tilde\rho_D(x,s)}{\partial x^2}&-(s+r)\tilde \rho_D(x,s)+\delta(x)\left[1+\tilde{j}(s)\right]=0,
\label{lap-M} \\
D\dfrac{\partial^2 \tilde\rho_R(x,s)}{\partial x^2}&-\dfrac{\partial [F(x)\tilde\rho_R(x,s)] }{\partial x}-s\tilde \rho_R(x,s)+r\tilde \rho_D(x,s)-\delta(x)\tilde{j}(s)=0 \label{lap-R},
\end{align}
where we have used the initial conditions $\rho_D(x,0)=\delta(x)$ and $\rho_R(x,0)=0$.
{Using the boundary conditions from Eq.~\eqref{rhoD-infinity}, the solution of Eq.~\eqref{lap-M} can be easily found as}
\begin{align}
\tilde \rho_D(x,s)=A(s)~\exp \left[-\sqrt{\frac{s+r}{D}}|x| \right],
~~\text{for}~~-\infty < x <\infty, \label{posmin-M}
\end{align}
where $A(s)$ is a $s$-dependent constant. {To compute $A(s)$, we integrate both sides of Eq. (\ref{lap-M}) over $[-\epsilon,\epsilon]$ around $x=0$ and finally take the $\epsilon \to 0$ limit. This provides us the following jump condition }
\begin{align}
D\left[\dfrac{\partial \tilde\rho^+_D(x,s)}{\partial x}\Big{|}_{x \to 0^+}-\dfrac{\partial \tilde\rho^-_D(x,s)}{\partial x}\Big{|}_{x \to 0^-}\right]=-[1+\tilde{j}(s)],
\label{mat-cond-1}
\end{align}
where the superscripts $^\pm$ indicate the solutions in the positive and negative sides of the origin.
Substituting $\tilde\rho^\pm_D(x,s)$ from Eq. \eqref{posmin-M} into Eq. \eqref{mat-cond-1}, we find 
\begin{align}
A(s)=\dfrac{1+\tilde j(s)}{2 \sqrt{D(r+s)}},
\label{con-A}
\end{align}
where $\tilde j(s)$ is still unknown and has to be determined from the return phase density as we will show now.    
Integrating both sides of Eq.~\eqref{lap-R} over a small region across the origin $x=0$, we arrive at the second discontinuity condition
\begin{align}
D\left[\dfrac{\partial \tilde\rho^+_R(x,s)}{\partial x}-\dfrac{\partial \tilde\rho^-_R(x,s)}{\partial x}\right]_{x=0}=\tilde{j}(s).
\label{current-LT}
\end{align}
Hence, using Eqs. (\ref{lap-R}) and (\ref{current-LT}), we can now solve for $\tilde{j}(s)$ which, in turn, gives $A(s)$, {although we still need to find $\tilde{\rho}_R^\pm(x,s)$.}


  The formalism presented so far provides a framework to compute the densities in Laplace space. Naturally, this framework can easily be applied to any set-up concerning an arbitrary confining potential  {$V(x)$}  exerted by the trap. Here, we consider the trap to be linear so that $V(x)=\lambda |x|$ {with $\lambda >0$}. This results in an external force $F(x)=-\lambda~\text{sgn}(x)$, where $\text{sgn}(x)=1$ for $x>0$,  $\text{sgn}(x)=-1$ for $x<0$ and equal to $0$ for $x=0$.
Using this expression of force in  Eq. \eqref{lap-R}, we write the following two differential equation for positive and negative $x$ regions
\begin{align}
D\dfrac{\partial^2 \tilde\rho^+_R(x,s)}{\partial x^2}+\lambda\dfrac{\partial [\tilde\rho^+_R(x,s)] }{\partial x}-s\tilde \rho^+_R(x,s)+rA(s)~e^{-\sqrt{\frac{s+r}{D}}x}&=0,~{\text{for}~x> 0,} \label{lap-pos-R}\\
D\dfrac{\partial^2 \tilde\rho^-_R(x,s)}{\partial x^2}-\lambda\dfrac{\partial [\tilde\rho^-_R(x,s)] }{\partial x}-s\tilde \rho^-_R(x,s)+rA(s)~e^{\sqrt{\frac{s+r}{D}}x}&=0,~{\text{for}~x< 0.}\label{lap-min-R}
\end{align}
Using the boundary conditions $\tilde\rho_R^{\pm}(\pm \infty,s)\to 0$ (from Eq. \eqref{rhoR-infinity}) and  $\tilde\rho_R^{\pm}(0,s)=0$ (from  Eq. \eqref{rhoR-abs}), we arrive at the full solution for the returning density
\begin{align}
\tilde\rho_R(x,s)=\dfrac{rA(s)}{\lambda\alpha_s-r}\left(e^{-\alpha_s |x|}-e^{-\mu_s |x|}\right),~~\text{for}~~-\infty < x <\infty,
\label{rho-R-LT-V}
\end{align}
with
\begin{align}
    \mu_s=\dfrac{\lambda+\sqrt{\lambda^2+4 s D}}{2D}~\text{and}~\alpha_s=\sqrt{\dfrac{s+r}{D}}~. 
\end{align}
We now substitute $\tilde{\rho}_R(x,s)$  and $A(s)$ from Eqs. (\ref{rho-R-LT-V}) and (\ref{con-A}), respectively, into Eq. (\ref{current-LT}) to find
\begin{align}
\tilde{j}(s)=\frac{r (\alpha_s -\mu_s )}{\mu_s  r-\alpha_s^2 \lambda }, \label{js}
\end{align}
which, in turn, results in
\begin{align}
A(s) = \frac{\lambda  \alpha_s-r}{\lambda  (r+2 s)-r \sqrt{4 D s+\lambda ^2}}{=\frac{\lambda  \sqrt{s+r}-r\sqrt{D}}{\sqrt{D}(\lambda  (r+2 s)-r \sqrt{4 D s+\lambda ^2})}}.
\label{A(s)}
\end{align}
Therefore, Eqs. (\ref{posmin-M}) and \eqref{rho-R-LT-V} along with the explicit expression for $A(s)$ provides us the full solutions for the master equations \eqref{rhoM-eqns} and \eqref{rhoR-eqns} in the Laplace space. The normalization condition for the joint density $\int_{-\infty}^{+\infty}~dx~\tilde\rho(x,s)=1/s$ also follows trivially.

Having fully solved the FP equations in Laplace space, the steady state density can be readily obtained by making use of the final value theorem which states $\rho^{\text{ss}}(x)=\lim_{s\to 0}[s \tilde{\rho}(x,s)]$. Applying this to  {Eqs.~\eqref{posmin-M} and \eqref{rho-R-LT-V} along with Eq.~\eqref{A(s)},} we find
\begin{align}
\rho_{D}^{\text{ss}}(x)&=\dfrac{\alpha_0 p_D }{2}~e^{-\alpha_0 |x|},\label{ss-1}\\
\rho_{R}^{\text{ss}}(x)&=\dfrac{r p_D}{2 (\lambda-\alpha_0 D)}\left( e^{-\alpha_0 |x|}-e^{-\frac{\lambda}{D} |x|}\right)
\label{ss-2},
\end{align}
{for $-\infty \leq x \leq \infty$} where $p_D=\lim_{t \to \infty}p_D(t)=\frac{\alpha_0 \lambda}{\alpha_0 \lambda+r}$ is the steady state probability for the system to be in the  {exploration} phase and  {recall that} $\alpha_0\equiv \alpha_{s=0}=\sqrt{\frac{r}{D}}$ is the typical inverse distance traveled by the particle  {in the  {exploration} phase} \cite{st-rt}. Note that the functions in Eqs.~\eqref{ss-1} and \eqref{ss-2} correspond to the right eigenfunctions of the Fokker-Planck operator with the largest eigenvalue which is zero. The steady state flux at the origin can be computed similarly and this gives
\begin{align}
\mathscr{J}_R(0,\infty)=\lim_{s\to 0}[s \tilde{j}(s)]=\dfrac{r\lambda}{\alpha_0 D+\lambda},
\end{align}
 {which is} consistent with $\mathscr{J}_R(0,\infty)=J_R(0^-,\infty)-J_R(0^+,\infty)$ as defined in Sec. \ref{model}.  For $\lambda \gg \alpha_0 D$,
one essentially reaches the limit of instantaneous resetting where flux due to resetting is $\mathscr{J}_R(0,\infty)=r$, as expected \cite{Restart1}. For more details on the properties of steady states we refer the readers to Ref.~\cite{st-rt}. In the following, we go beyond steady states and study time dependent solutions of the FP equations in \eqref{rhoM-eqns} and \eqref{rhoR-eqns} to understand the approach to the steady state. First, we present exact solutions for the time dependent propagators in the next section.

\section{Time dependent density $\rho(x,t)$}
\label{pb-t}
The position distribution functions in time domain can be obtained by performing the inverse Laplace transform so that
\begin{align}
\rho_{D,R}(x,t)=\dfrac{1}{2 \pi i}\int_{\Gamma-i\infty}^{\Gamma+i\infty}~ds~e^{st}~\tilde{\rho}_{D,R}(x,s)\label{ILT},
\end{align}
{which is expressed in terms of the Bromwich integral \cite{wong}}. {Here,} the integration is done along the imaginary axis
(vertical contour through $\text{Re}(s)=\Gamma$) in the complex-$s$ plane such that all the singularities lie on the left of it. We first recall the densities in the exploration and return phases from Eqs. (\ref{posmin-M}) and \eqref{rho-R-LT-V} 
\begin{align}
\tilde\rho_D(x,s)&=\dfrac{(\lambda \sqrt{s+r}-r\sqrt{D})[\lambda(r+2 s)+2 r\sqrt{D} \sqrt{s+\beta}]}{4 \lambda^2\sqrt{D} s(s-\gamma)}~e^{-\sqrt{s+r}|z|},\label{rem-1}\\
\tilde\rho_R(x,s)&=\dfrac{r[\lambda(r+2 s)+2 r\sqrt{D} \sqrt{s+\beta}]}{4 \lambda^2 s(s-\gamma)}~\left(e^{-\sqrt{s+r}|z|}-e^{-\sqrt{\beta} |z|}~e^{-\sqrt{s+\beta}|z|}\right)\label{rem-2},
\end{align}
where we have used Eq.~\eqref{A(s)} and defined
\bea
{
z=\frac{x}{\sqrt{D}},~~\beta=\frac{\lambda^2}{4 D},~~\gamma=\frac{r^2}{4\beta}-r.} 
\label{z(x)}
\eea
 It is easy to see that $\tilde\rho_{D,R}(x,s)$ have three singularities for $r\neq \beta$: one pole at $s=0$ and two branch points at $s=-r$ and $s=-\beta$. Naturally when $r=\beta$, we have one pole at $s=0$ and one branch point at $s=-r=-\beta$. Moreover, substituting $s$ by $1/u$, one can see that there is one more branch point at infinity which we choose to be at $s=-\infty$ for both cases. Looking at the expressions in (\ref{rem-1})-(\ref{rem-2}) at a first glance, it seems that there is a pole also at $s=\gamma$ for both $\tilde{\rho}_D$ and $\tilde{\rho}_R$, however it turns out to be a removable singularity.  {To see this, we first note that $s-\gamma=\frac{D}{\lambda^2}(\lambda\alpha_s-r)(\lambda \alpha_s+r)$.  
 The factor $(\lambda\alpha_s-r)$ gets cancelled with the first factor term in the numerator in Eq.~\eqref{rem-1}. On the other hand, for $\tilde{\rho}_R(x,s)$, the cancellation happens differently for $r>\beta$ and $r < \beta$. In the former case, 
 note from Eq.~\eqref{rem-2} that $e^{-\sqrt{s+r}|z|}-e^{-\sqrt{\beta} |z|}~e^{-\sqrt{s+\beta}|z|}={(s-\gamma) g_1(\gamma) + O((s-\gamma)^2)}$, which essentially eliminates the factor $s-\gamma$ in the denominator and thus removing the pole at $s=\gamma$.
In the latter case, this cancellation is done by the factor term in the numerator in Eq. \eqref{rem-2}. }{Taking into account the structure of these singularities in details, we now proceed to perform the integrals in Eq.~\eqref{ILT}} using Bromwich contours {(see  Fig.~\ref{fig:BC}) for $r\neq \beta$ and $r=\beta$ cases separately in the next subsections.}

It is worth noticing that the parameters $r$ and $\beta$ are associated to two different time scales in the system. 
The first time scale $\tau_{res} = r^{-1}$ comes from the resetting events which provides the relaxation time for the density to reach steady state in case of instantaneous resetting \cite{Review}. On the other hand, $\beta = \frac{\lambda^2}{4D}$ provides the relaxation time scale $\tau_{pot}=\beta^{-1}$ for the particle to reach steady state in absence of any resetting events. In presence of both, these two times scales compete and naturally we get the following two cases:  $r=\beta$ and $r\neq \beta$. In the next sections we will observe relaxation terms with prefactors $e^{-rt}$ and $e^{-\beta t}$ appearing naturally in the expressions of
in $\rho_{D,R}(x,t)$.

\subsection{Case 1: $r=\beta$}
In this case, the branch points $s=-r$ and $s=-\beta$ coalesce with each other, and thus, we now have one pole at $s=0$ and two branch points at $s=-r$ and  $s=-\infty$ (hence one branch cut). We construct a contour as shown in Fig.~\ref{fig:BC} and perform the integral by considering the contributions from the branch {cuts} and the pole. 
We follow the steps described in \ref{sec:bc-int} and collect the contributions from the pole and the branch cuts.
After making some simplifications we get the following expressions for the densities in the exploration and return phases:
\begin{align}
&\rho_{D}(x,t)=\rho^{ss}_{D}(x)
+\frac{1}{4 \pi  \lambda ^2}\int_0^\infty~dy~\frac{e^{-t(r+y)}}{(r+y) (r+\gamma+y)}\bigg[\lambda  r^2 \sin \left(\sqrt{y} \left| z\right| \right) \nonumber\\
&~~~~~~~~~~~~~~~~~~~~~~~~~~~~~~~~+~\sqrt{\frac{y}{D}}\left(2 r^2 D +\lambda ^2 r+2 \lambda ^2 y\right)\cos \left(\sqrt{y} \left| z\right| \right) \bigg],\label{reqbeta-m}\\
&\rho_{R}(x,t)=\rho^{ss}_{R}(x)-\frac{r}{4 \pi  \lambda ^2}\int_0^{\infty}~dy~\frac{e^{-t(r +y)}(1-e^{-\sqrt{r}|z|})}{(r +y) (r +\gamma+y)}\bigg[2 r \sqrt{D y} \cos \left(\sqrt{y} \left| z\right| \right)\nonumber\\
&~~~~~~~~~~~~~~~~~~~~~~~~~~~~~~~~~~~~~~~~~~~
~~~~~~~~~~~~
+\lambda  (r+2 y) \sin \left(\sqrt{y} \left| z\right| \right)\bigg]\label{reqbeta-r},
\end{align}
where $z=x/\sqrt{D}$ and steady state densities $\rho^{\rm ss}_{D,R}(x,t)$  are given in Eqs.~\eqref{ss-1} and \eqref{ss-2}.
{We verify {these expressions} against numerical simulations in Fig.~\ref{fig:num-sim}. Although the results presented above
are exact, they} do not provide much insights. Moreover, it is difficult to quantify from these expressions how the densities relax to their corresponding steady states. In Sec. \ref{relaxation}, we will analyze these expressions to understand the relaxation behaviour.

\subsection{Case 2: $r\neq \beta$}
\begin{figure}[]
  \begin{center}
    \includegraphics[width=5cm]{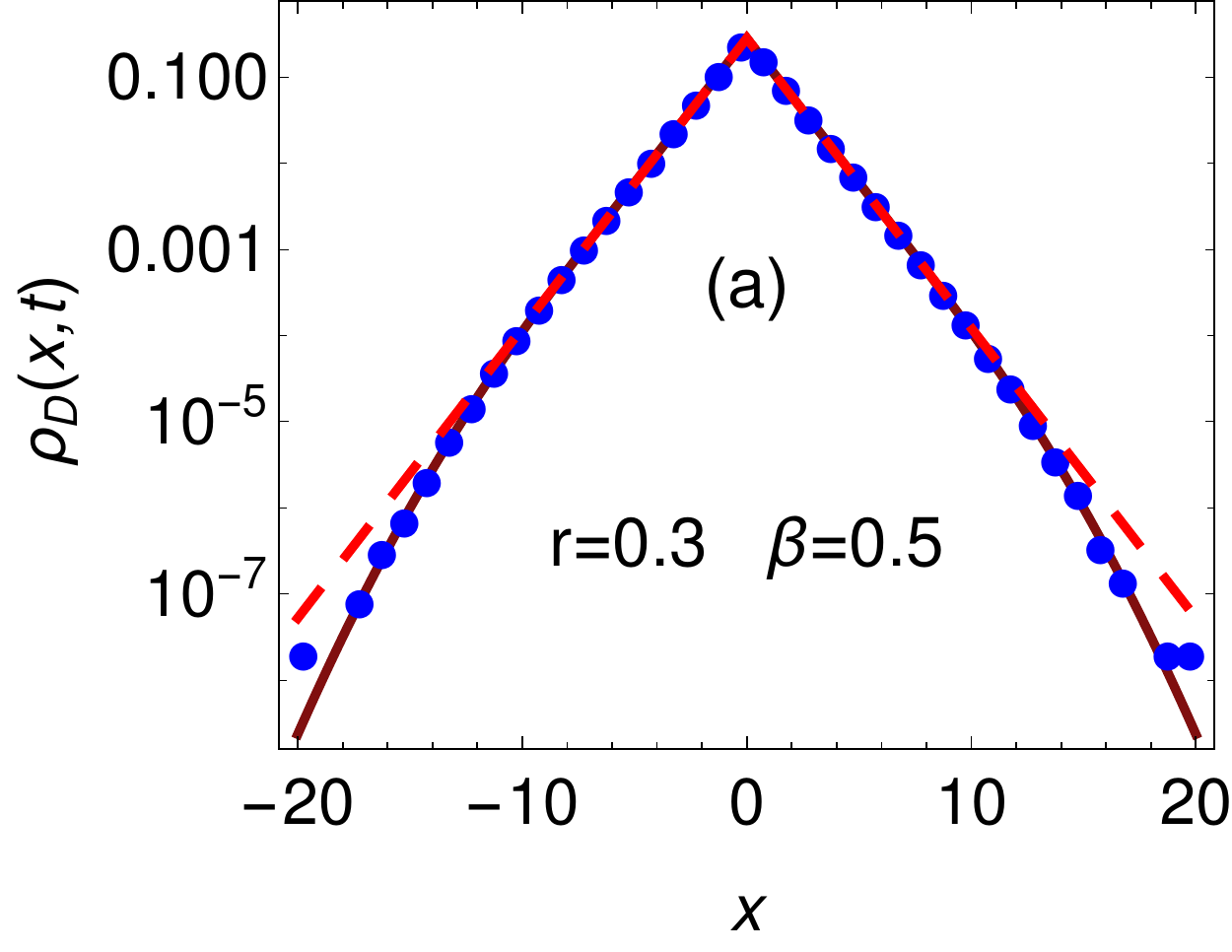}~~
    \includegraphics[width=5cm]{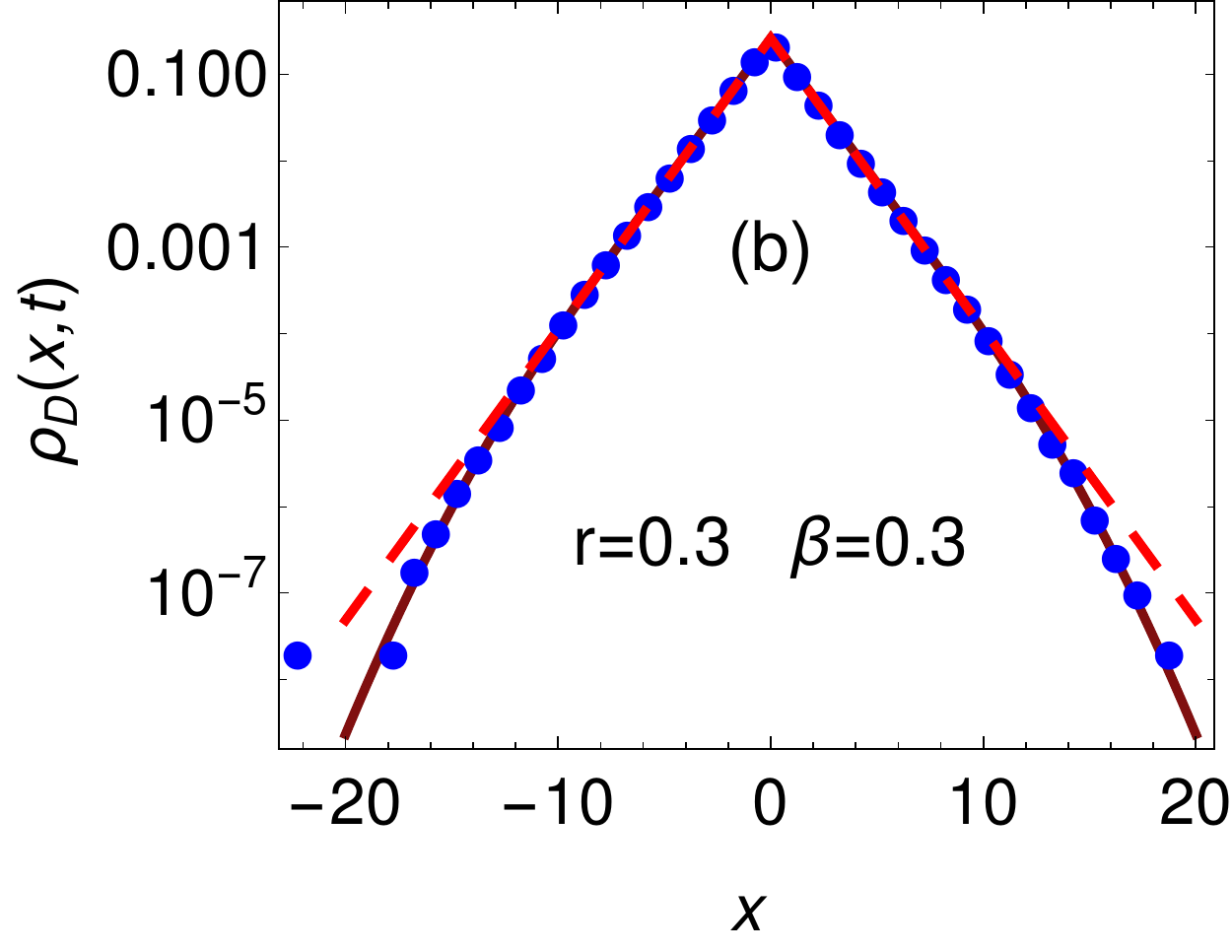}~~
    \includegraphics[width=5cm]{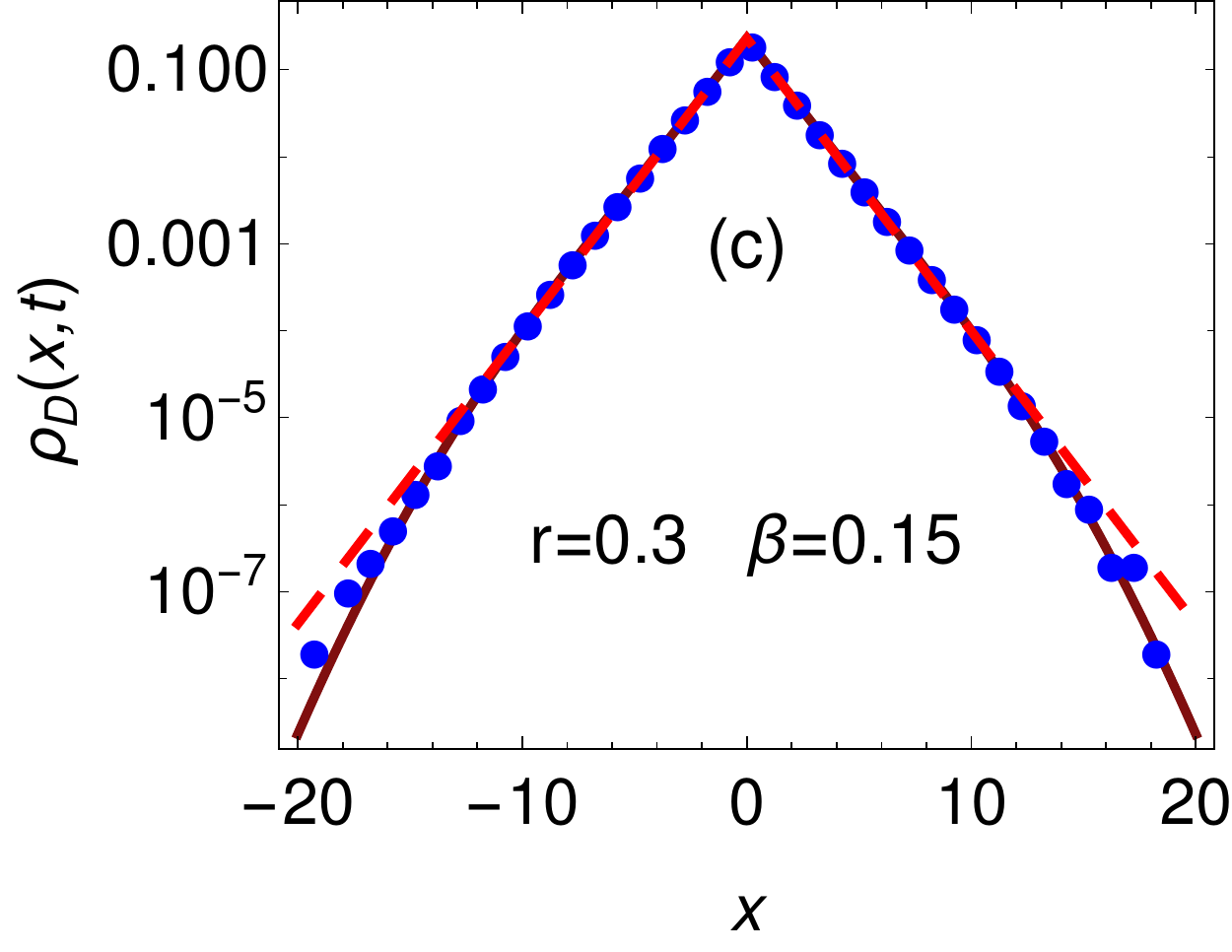}\\
    \includegraphics[width=5cm]{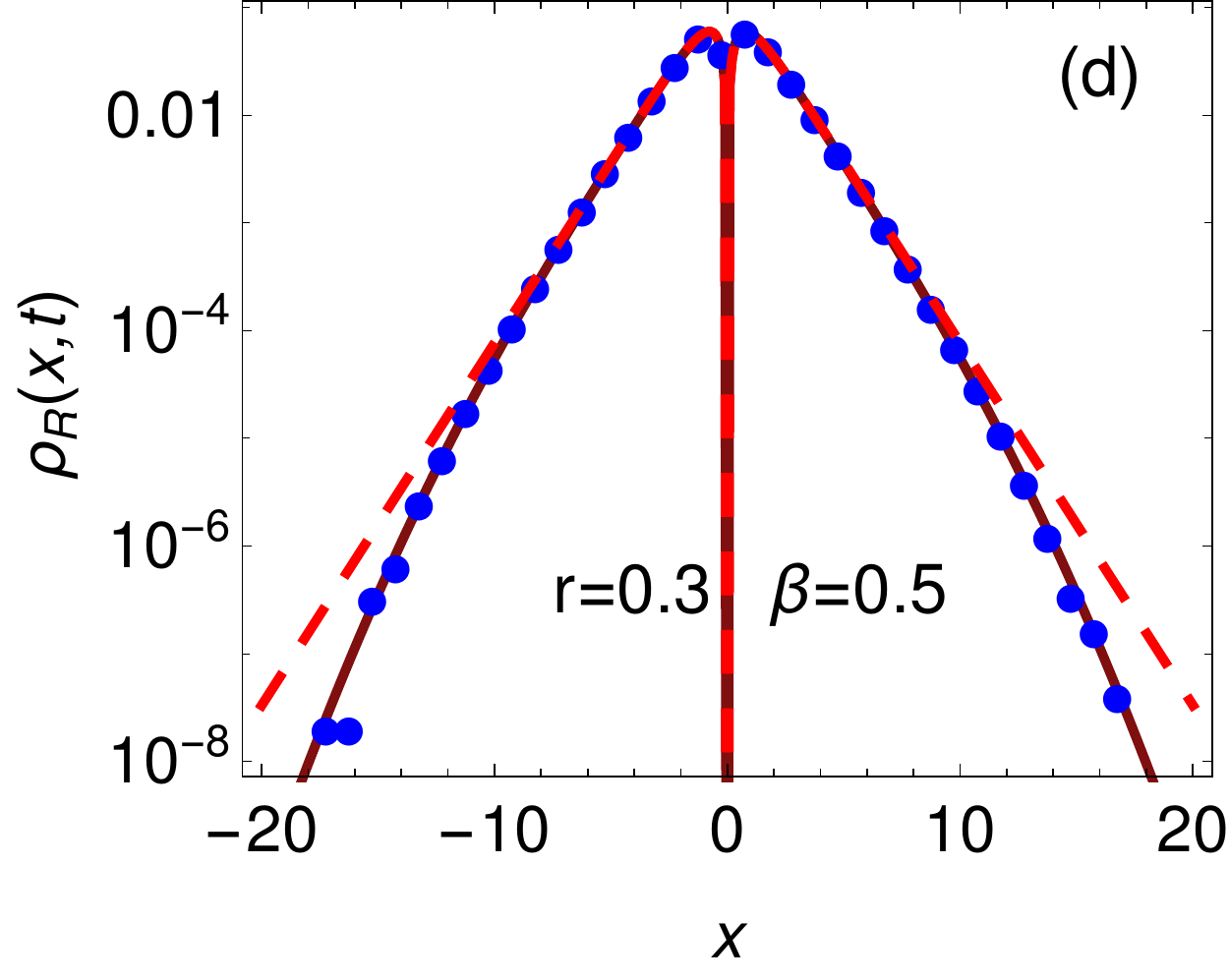}~~
    \includegraphics[width=5cm]{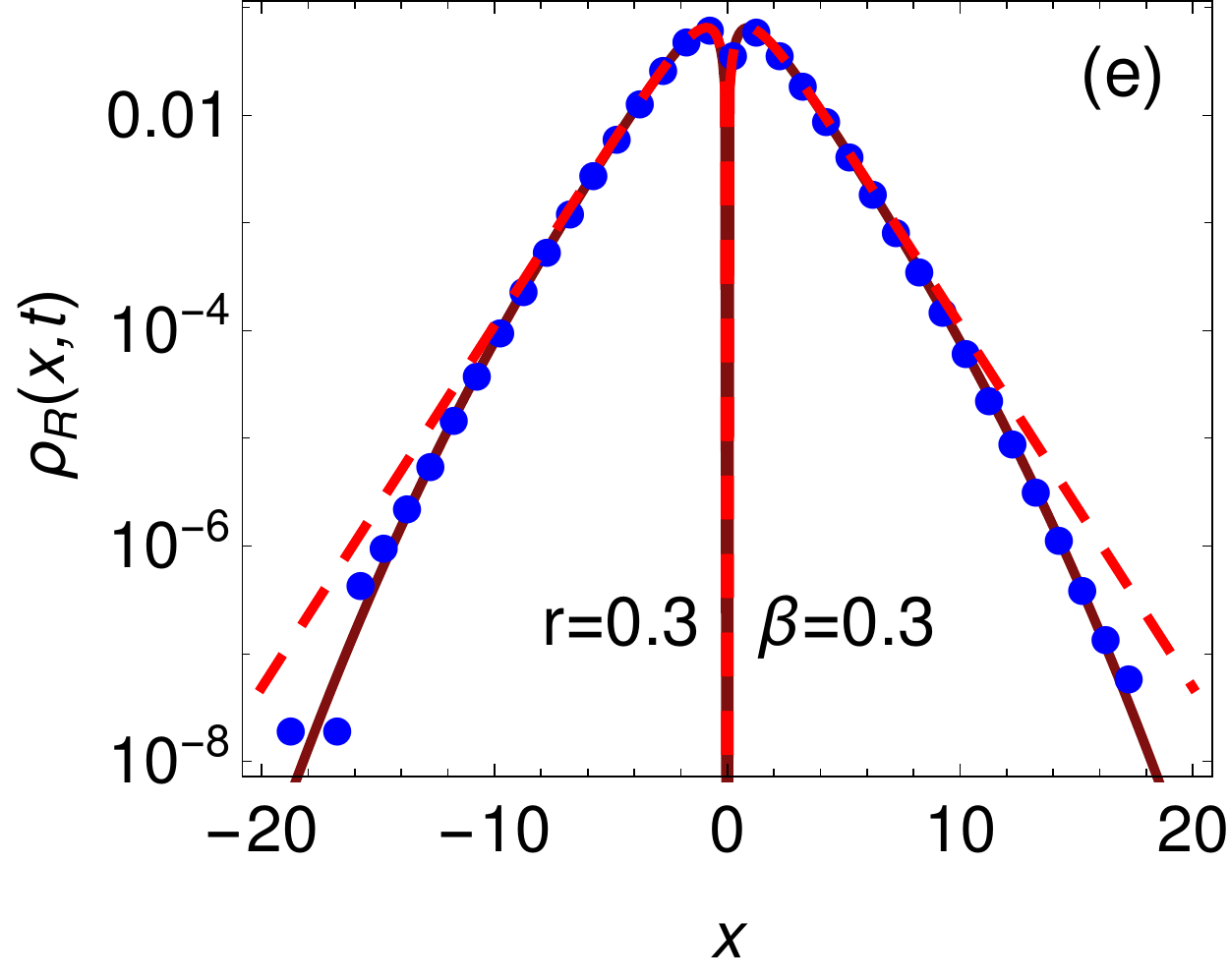}~~
    \includegraphics[width=5cm]{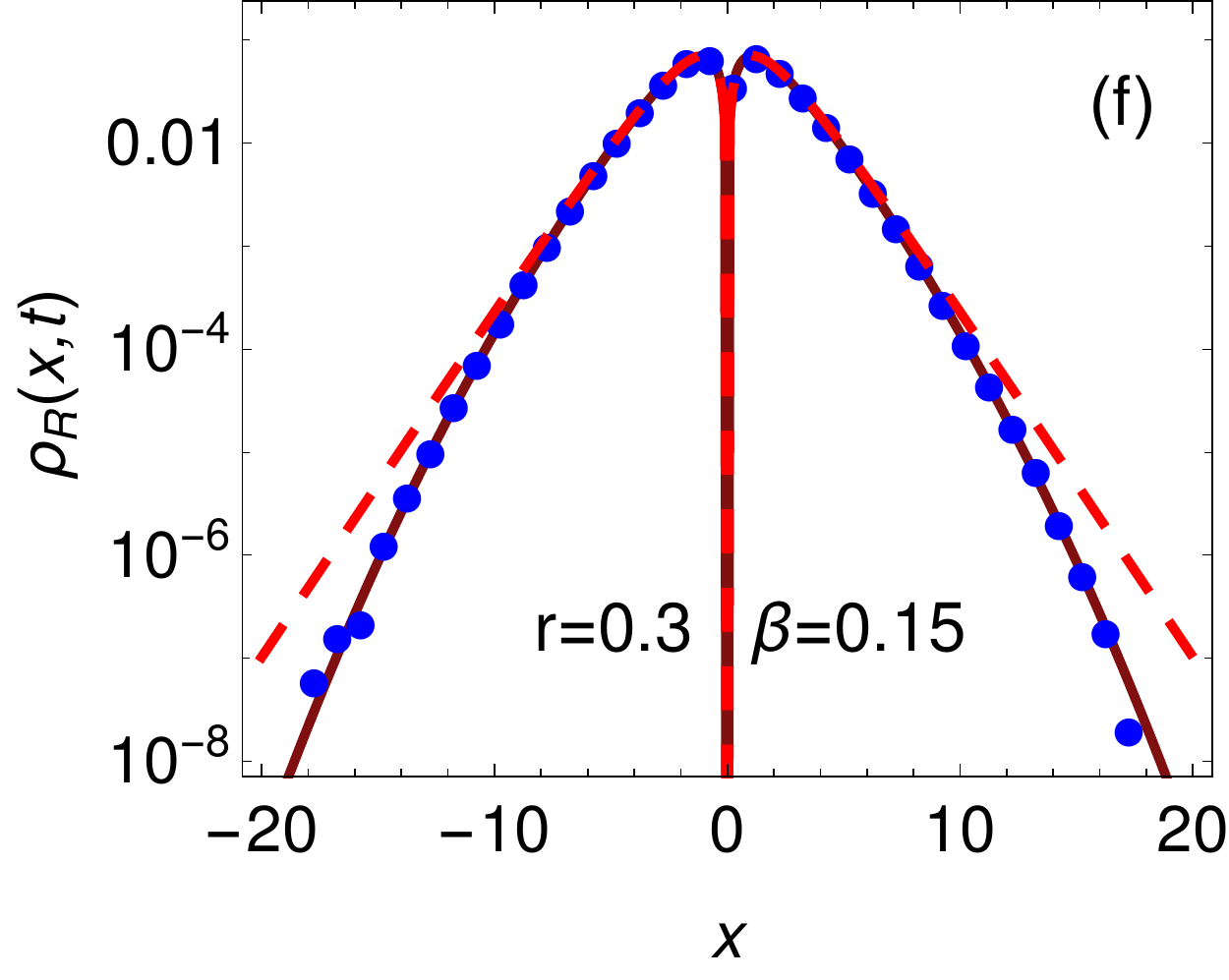}\\
    \includegraphics[width=5cm]{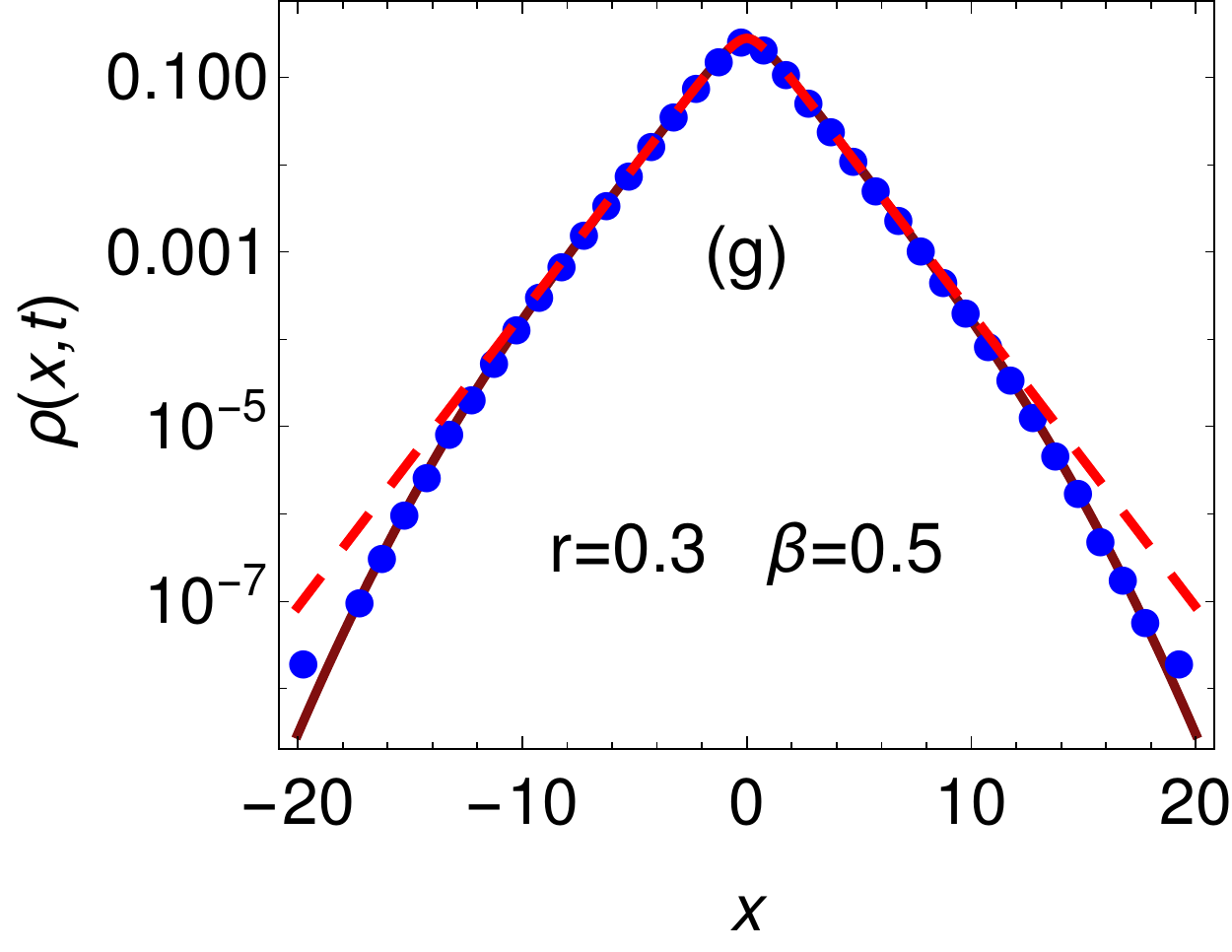}~~
    \includegraphics[width=5cm]{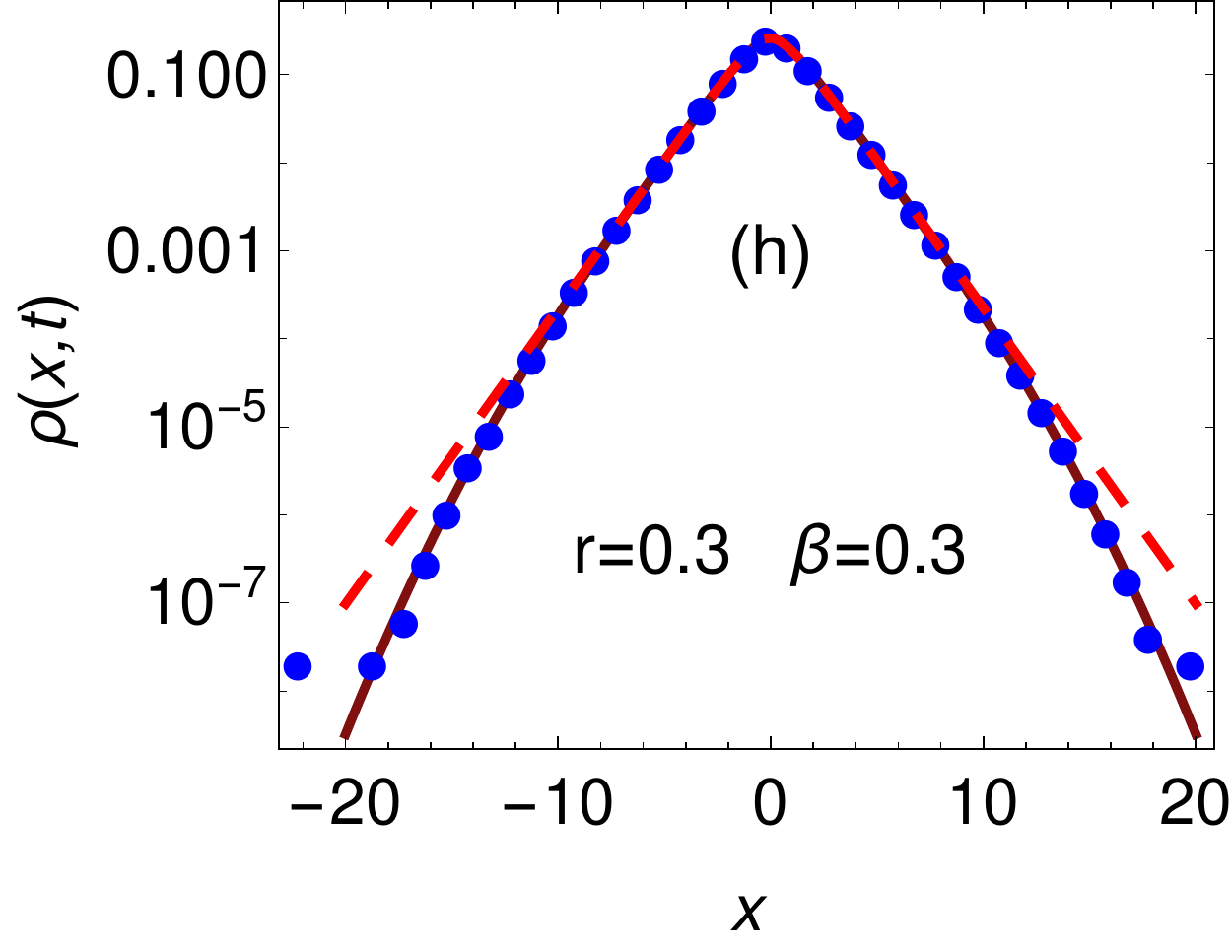}~~
    \includegraphics[width=5cm]{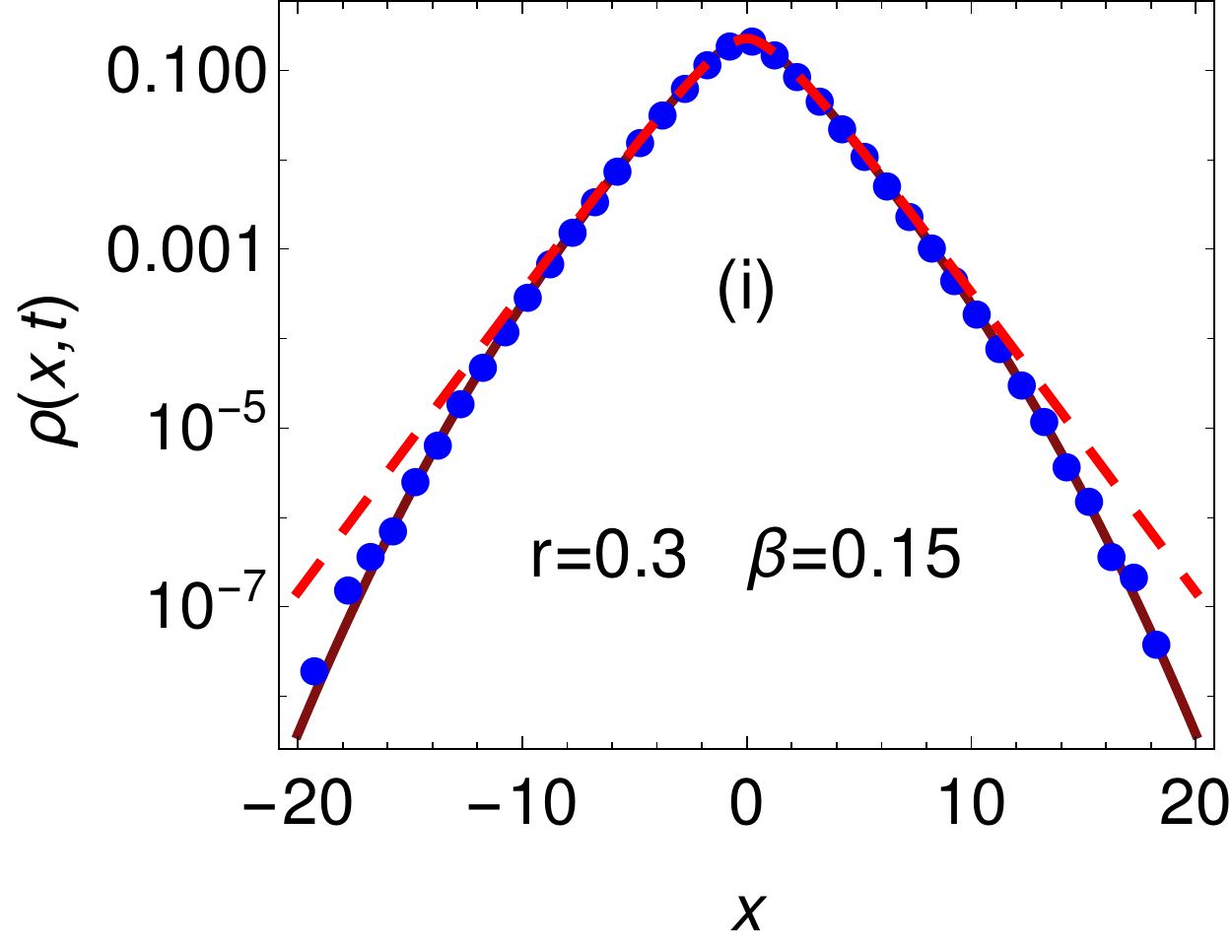}\\
    \caption{Comparison of analytical results for density functions $\rho_D(x,t)$, $\rho_R(x,t)$ and $\rho(x,t)=\rho_D(x,t)+\rho_R(x,t)$, with the same obtained from numerical simulation. First, second, and third-column, respectively, correspond to three different cases: $r<\beta$, $r=\beta$, $r>\beta$. Solid curves are the analytical results given in Eqs.~(\ref{rlbeta-m}-\ref{part-2}) whereas the circles are obtained from numerical simulation for $10^8$ realizations. In each plot, dashed curves indicates the steady state densities given in Eqs.~\eqref{ss-1} and \eqref{ss-2}. Common parameters used in these plots are  $D=0.5$, $dt=10^{-3}$, $t=15$.}
    \label{fig:num-sim}
  \end{center}
\end{figure}

In this case, there are three branch points: $s=-r, s=-\beta, s=-\infty$, and one pole at $s=0$.  We refer to \ref{sec:bc-int} for {details}  of these Laplace inversions 
{to obtain expressions of the densities. Since these expressions are {quite lengthy}, we prefer to present them 
in the Appendix A (see Eqs.~(\ref{rlbeta-m}-\ref{part-2})).}
In  Fig. \ref{fig:num-sim}, we  compare the expressions for the densities in the exploration and the return phases given in  Eqs.~(\ref{rlbeta-m}-\ref{part-2}), against numerical simulations for both $r>\beta$ and $r<\beta$.  We observe excellent agreement between the two in each case.

\section{Probability to be in the exploration and return phases}
\label{PD-PR}
Before we get into the relaxation of the  density $\rho(x,t)$, it is imperative to study simple quantities to get some insights. To this end, in this section, we study the probability for the particle to be in the exploration phase. This quantity was formally defined in Sec.~\ref{model}. We recall here for brevity
\bea
p_D(t)=\int^{+\infty}_{-\infty}~dx~ \rho_D(x,t).
\label{p-D-t}
\eea
Since the particle starts in the exploration phase, we have $p_D(0)=1$ and then the probability decreases with time and finally saturates to a steady state value ($<1$) at large time.
In this section, we obtain an exact expression for $p_D(t)$. For simplicity, we will consider the case when $r=\beta$ but the other cases also can be computed 
following the same procedure. To begin with, we integrate $\tilde\rho_D(x,s)$ given in Eq.~\eqref{rem-1} over the entire real space, and we obtain 
\begin{align}
    \tilde p_D(s)=\dfrac{2\sqrt{s+r}-\sqrt{r}}{(2s+r)\sqrt{s+r}-\sqrt{r}(s+r)},
    \label{pds}
\end{align}
where we have used the condition $\lambda=\sqrt{4 D r}$ that comes from $r=\beta$ limit.
The probability $p_D(t)$ can be obtained using the inverse Laplace transform 
\begin{align}
    p_D(t)=\dfrac{1}{2\pi i}\int_{\Gamma-i\infty}^{\Gamma+\infty}~ds~e^{st}~\tilde p_D(s).
\end{align}
Clearly, we can see from Eq.~\eqref{pds} that the integrand of the above equation has one pole at $s=0$ and one branch point at $s=-r$. Therefore, we consider the contour shown in Fig.~\ref{fig:BC} with the branch points $s_b^{(1)}$ and $s_b^{(2)}$ coalescing at $s=-r$. Following similar steps of calculation as shown in \ref{sec:bc-int}, we finally get
\begin{align}
    p_D(t)=\dfrac{2}{3}+\dfrac{r^{3/2}e^{-rt}}{\pi}\int_0^\infty~dy~\dfrac{e^{-y t}}{\sqrt{y}(r+y)(r+4 y)}.\label{eq-pdt}
\end{align}
where the integral can be computed easily. The full expression for $p_D(t)$ is given by
\begin{align}
   p_D(t)=\dfrac{2}{3}+ \frac{1}{3}\left[2 e^{-\frac{3r t}{4}} \text{erfc}\left(\frac{\sqrt{r t}}{2}\right)- \text{erfc}\left(\sqrt{r t}\right)\right],\label{eq-pdt}
\end{align}
where $\text{erfc}(u)=\frac{2}{\sqrt{\pi}}\int_u^\infty~dt~e^{-t^2}$ is the complementary error function. The probability to find the particle in the return phase at time $t$ is then simply given by $p_R(t)=1-p_D(t)$.

\begin{figure}
    \centering
    \includegraphics[width=8cm]{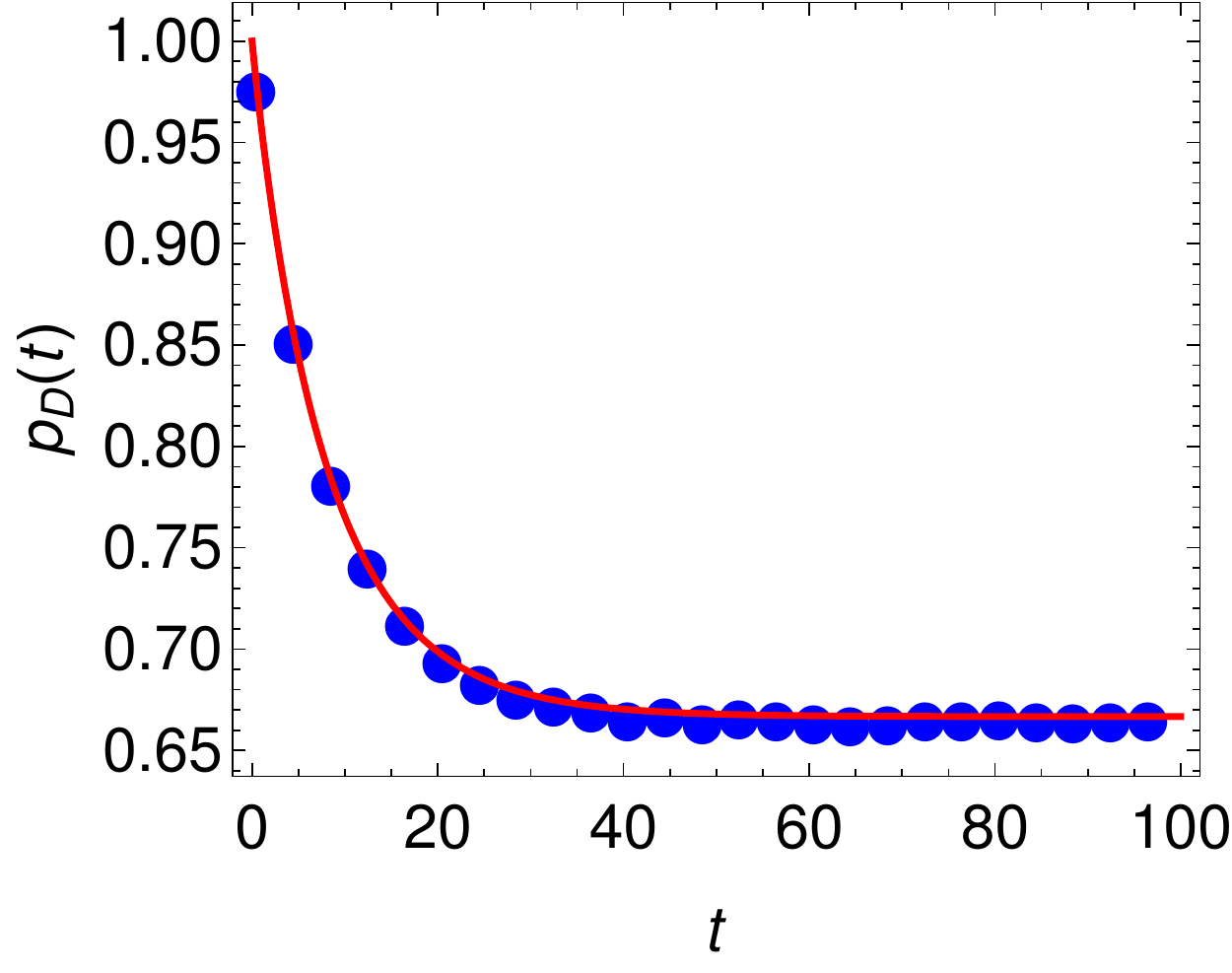}
    \caption{Probability of the system in the exploration phase with time. Solid line is the analytical result shown in Eq.~\eqref{eq-pdt} while circles are obtained using numerical simulation for $10^5$ realizations. The parameters for the simulation are $r=\beta=0.1$, $D=0.5$, and $dt=10^{-3}$.}
    \label{pdt}
\end{figure}

At large time, we see that $p_D(t)$, relaxes exponentially as $e^{-rt}$ to its steady state value $2/3$ for $r=\beta$. Note here that $p_D(\infty)$ is nothing but the fraction of time spent in the exploration phase in steady state such that
\bea
p_D(\infty)=\frac{1/r}{1/r+\langle \overline{\tau}(x) \rangle},
\label{pm-exact}
\eea
where $1/r$ and $\langle \overline{\tau}(x) \rangle$ are the mean times spent in the exploration and return phases respectively. In \cite{st-rt}, it was shown that $\langle \overline{\tau}(x) \rangle=\frac{1}{\alpha \lambda}$, which upon plugging to Eq. \eref{pm-exact} yields the value as expected.
In Fig.~\ref{pdt}, we show the comparison of the analytical result Eq.  \eqref{eq-pdt} with the numerical simulation for $r=\beta=0.1$ and they show a very good agreement.
Setting the stage, we now delve deeper to understand the relaxation of the probability densities. This is done in the next section.

\section{Relaxation to the steady state}
\label{relaxation}
Presence of resetting affects the stochastic dynamics of a particle in many different ways. For example, a free Brownian particle under resetting mechanism reaches a steady state which in absence of resetting is described by a Gaussian propagator with variance growing linearly with time. Moreover, the relaxation properties to the steady state is also unusual compared to the equilibration of the same particle in a confining potential in the absence of resetting. In the context of instantaneous resetting of a free Brownian particle, it has been shown that at a given large time $t$ the probability density $\rho(x,t)$ as a function of $x$ exhibits two markedly different phases: 
an inner core which has already reached the steady state and an outer core which is still explicitly time dependent \cite{Review,transport1}. The boundaries that separate these two regions grow linearly with time. Physically, these two regions are created by two groups of trajectories: ones that have  encountered zero (almost zero) resetting events up to the observation time $t$ and the ones that have undergone many resetting events. The former group contributes to the outer region and the latter group contributes to the inner core region of the density function. Such `cone spreading' relaxation with travelling fronts is studied and observed in few other stochastic dynamics with instantaneous resetting e.g., underdamped Brownian particle \cite{underdamped}, interface dynamics \cite{transport1}, random acceleration process \cite{RAP} and resetting occurring with time dependent rate~\cite{Pal-time-dep}.

Notably, in all these resetting systems the relaxation properties are studied using the renewal approach. However, in many problems it might be difficult to follow the renewal approach analytically. In such cases one would require to rely upon the master equation approach where the relaxation is usually described through spectral decomposition. In this section, we demonstrate how the above mentioned `cone spreading'/`traveling boundary' type relaxation behavior can be understood also from the master equation approach. 
We first discuss the limit $\lambda \to \infty$ (more precisely, $\lambda \gg \alpha_0 D$) which corresponds to the well studied instantaneous resetting scenario and recover the existing results. Subsequently, we extend our analysis to the non-instantaneous resetting case, (i.e., finite $\lambda$) and provide new insights that emerge due to the finite duration of return events.

\subsection{Instantaneous resetting}
\label{inst-reset}
We start with the case of instantaneous resetting. Taking the limit $\lambda \to \infty$ in Eqs. \eqref{rem-1} and \eqref{rem-2}, we find
\begin{align}
\tilde\rho_R(x,s)=&0, \\
\tilde \rho_D(x,s)=&\dfrac{1}{2s}\sqrt{\dfrac{s+r}{D}}~e^{-\sqrt{s+r}|z|} \label{rhoD(z,s)-instant}.
\end{align}
Since the return events are instantaneous, $\rho_D(x,t)$ is essentially the total density function. Hence, in the following we drop its subscript $D$.  To get $\rho(x,t)$ we perform the inverse Laplace transform using the following Bromwich integral
\begin{align}
\rho(x,t)=\dfrac{1}{2 \pi i}\int_{\Gamma-i\infty}^{\Gamma+i\infty}~ds~e^{st}~\tilde \rho(x,s),\label{brom}
\end{align}
where $\Gamma$ is the vertical contour passing through $\Gamma=\mathrm{Re}(s)$ of the complex $s$-plane (see Fig.~\ref{fig:BC} in \ref{sec:bc-int}).
We {notice} that $\tilde \rho(x,s)$ in Eq.~\eqref{rhoD(z,s)-instant} has two branch points at $s=-r$ and $s=-\infty$, and a pole at $s=0$ {and they all}  lie to the left of the vertical contour $\Gamma$. 
{Following the steps sketched in \ref{sec:bc-int},}
we finally obtain
\begin{align}
\rho(x,t)=\frac{\alpha_0}{2}~e^{-\sqrt{r}|z|}+\frac{e^{-r t}}{2\pi \sqrt{D}}\int_0^\infty~dy~\dfrac{\sqrt{y}~e^{-y t}}{y+r}\cos \left(|z|\sqrt{y}\right),~\text{with}~z=\frac{x}{\sqrt{D}}, 
\label{inst-ex}
\end{align}
where the first term is the contribution from the pole and the second term gets contribution from the branch {cut $(-\infty,-r]$.} Further, we make a change of variable $y=w^2$ and extend the limit of integration on the entire real line by writing $\cos(w)=\frac{e^{i w}-e^{-i w}}{2}$ {which} yields
\begin{align}
\rho(x,t)&=\sqrt{\frac{r}{D}}\frac{e^{-\sqrt{r}|z|}}{2}+\frac{e^{-r t}}{2\pi \sqrt{D}} \mathcal{K}_{\sqrt{r}}^{\mathcal{A}^{(0)}}(x,t),~~~\text{where,}\label{pdf} \\
\mathcal{K}_{a}^{\mathcal{A}}(x,t) &=  \int_{-\infty}^{+\infty}~dw~\frac{\mathcal{A}(w)}{w-ia}~e^{-tw^2+iw|z|},~~~{\text{with}~z=\frac{x}{\sqrt{D}}}~~\text{and},
\label{u-part} \\
\mathcal{A}^{(0)}(w)&=\frac{w^2}{w+i\sqrt{r}}. \label{mcalA(w)}
\end{align}
Note that Eq. \eqref{pdf} provides the exact density function at all time. Also note that the first term in Eq.~\eqref{pdf} is the steady state density $\rho^{\rm ss}(x)$ of the Brownian particle under instantaneous resetting \cite{Restart1}. The second term provides the relaxation to this steady state. To understand this relaxation better and in detail, we now analyze the integral in Eq.~\eqref{u-part} at large $t$. For that we extend this integral in the complex $w$ plane and perform the integral on the closed contour $\mathcal{C}$ as shown in Fig.~\ref{fig:contour-C}. Performing the saddle point integral as sketched in \ref{AppendixB}, we evaluate $ \mathcal{K}_{\sqrt{r}}^{\mathcal{A}}(x,t)$ and obtain the following approximate expression of the second term in Eq.~\eqref{pdf} for large $t$:
\begin{align}
\frac{e^{-r t}}{2\pi \sqrt{D}} \mathcal{K}_{\sqrt{r}}^{\mathcal{A}^{(0)}}(x,t) &= -\frac{e^{-\sqrt{r}|z|}}{2}\sqrt{\frac{r}{D}}\Theta(|z|-\sqrt{4r}t )+\frac{e^{-rt-\frac{z^2}{{4}t}}}{2\pi \sqrt{D}} ~K_z^{\mathcal{A}^{(0)}}\left(\sqrt{r}-\frac{|z|}{2t},t\right),\label{2nd-term-ins} \\
 \text{where,}  ~
 K_z^{\mathcal{A}^{(0)}}(b,t) &\approx \mathcal{A}^{(0)}\left(ib+i\frac{|z|}{2t}\right)e^{tb^2} \left[ i\pi~\text{sgn}(b) ~\text{erfc}(|b|\sqrt{t})\right]+\psi(0)\sqrt{\frac{\pi}{t}}~,
    \label{J-b-final-mt}
\end{align}
with $\psi(0)=\frac{i}{b}\left[ \mathcal{A}^{(0)}\left(i{|z|}/{2t}\right)-\mathcal{A}^{(0)}\left(ib+i{|z|}/{2t}\right)  \right]$.
The first term in Eq.~\eqref{2nd-term-ins} appears for $|z| > \sqrt{4r}t$ and interestingly, it is exactly same as  the first term in Eq.~\eqref{pdf} but with opposite sign. As a result the steady state form for $\rho(x,t)$ serves as the dominant contribution for $|x|\leq \sqrt{4Dr}t$ and thus creating a core region which expands in time with a speed $\sqrt{4 D r}$ on both sides of $x=0$ \cite{transport1}. Outside this region, $\rho(x,t)$ is explicitly time dependent.
Performing further simplifications and keeping the dominant order terms in the large $t$ limit, we get the following simpler and more explicit asymptotic form for the density as 
\begin{align}
\begin{split}
\rho(x,t)\approx \frac{1}{2}\sqrt{\frac{r}{D}}e^{-\sqrt{r}|z|}&\Theta(\sqrt{4r}t -|z|) + \dfrac{e^{-r t} e^{-z^2/4t}}{ \sqrt{4 \pi D t}} \left[ \frac{z^2}{z^2-4rt^2}-\frac{\sqrt{r}t}{|z|-\sqrt{4r}t} \right]  \\
    &+\frac{\alpha_0}{4} \text{sgn}(|z|-2\sqrt{r}t)~e^{-\sqrt{r}|z|}~\text{erfc}\left(\frac{||z|-2\sqrt{r}t|}{\sqrt{4t}}  \right),
\end{split}
\label{rho_z-t-inst-aprx}
\end{align}
where recall $z=x/\sqrt{D}$. This expression clearly describes the `cone spreading' relaxation behaviour with travelling fronts, first observed using renewal approach in Ref.~\cite{transport1}. In this reference the relaxation to the steady state is interpreted as a dynamical transition occurring in the system when observed at a fixed spatial point. The transition was further characterised by non-analytic properties of an appropriate large deviation function (LDF) \cite{transport1}.
{Looking at the large $t$ asymptotic of the different terms in  Eq.~\eqref{rho_z-t-inst-aprx} it is easy to see that the density in the large $t$ limit satisfies a large deviation principle {(for a review on large deviation principle, see Ref. \cite{ldf-rep})} 
\begin{align}
\rho(x,t) &\asymp e^{-t\Phi(x/t)},~\textit{i.e.}~\Phi\left(v=\frac{x}{t}\right)=-\lim_{t \to \infty} \frac{\ln \rho(x,t)}{t}~\text{where},~ 
\label{LD-form} \\
\Phi\left(v\right)&= 
\begin{cases}
\sqrt{\frac{r}{D}} |v|,~&\text{for}~|v|< v^* \\
r+\frac{v^2}{4D},~&\text{for}~|v|\geq v^* ,
\end{cases}
~\text{with}~v^*=\sqrt{4 D r}.
\label{LD-ins}
\end{align}
The non-analytic structure of the large deviation function $\Phi(v)$ at $v=v^*$ is interpreted as a dynamical transition. This transition mentioned above is manifested as follows: 
at a fixed spatial point $x$, the LDF is quadratic at short time when $|v|>v^*$. On the other hand, as time increases, $v$ decreases and eventually 
crosses $v^*$ from above. In this case, the LDF changes from the quadratic to a linear form as in \cite{transport1}.}


While renewal approach clearly depicts the physical mechanism behind this unusual relaxation behaviour,  our analysis reveals that mathematically this transition happens exactly when the saddle point hits the pole, i.e., when $w^*=w_+$ which gives $|z^*|=t \sqrt{4r}$ i.e., $|x^*|=t \sqrt{4Dr}$ [see Fig.~\ref{fig:contour-C}]. In a broader sense, our approach identifies a connection between the structure of the singularities in propagator in Laplace space and the relaxation behaviour of the particle.

\subsection{Non-instantaneous resetting}
{We now turn our attention to the non-instantaneous resetting. Recalling from Sec.~\ref{pb-t}, there are three different regimes $r<\beta$, $r=\beta$ and  $r>\beta$ for which the exact time dependent densities were obtained in Sec.~\ref{pb-t}.} 
In the following, we discuss the relaxation properties for  these cases separately using the saddle point method applied in the previous section. In each case, we will  write the total density $\rho(x,t)=\rho_D(x,t)+\rho_R(x,t)$ in the form of Eq.~\eqref{pdf} i.e., the steady state part $\rho^{\rm ss}(x)$ plus relaxation parts which are in the form of integral in Eq.~\eqref{u-part}. 

\subsubsection{\textbf{Case I ($r=\beta$):~}}
\label{sec:case-1}
In this case, the densities in the exploration and return phases were already obtained in Eqs.~\eqref{reqbeta-m} and \eqref{reqbeta-r}. Adding these two densities and simplifying, we find that the we total density $\rho(x,t)$ can be written in the form of Eq.~\eqref{pdf} as 
\begin{align}
    \rho(x,t)&=\rho^{ss}(x)+\frac{e^{-r t}}{8\pi\sqrt{D}} \mathcal{K}_{\sqrt{r}}^{\mathcal{A}^{(1)}}(x,t),~\text{with} 
    \label{r-eq-beta} \\
  \mathcal{A}^{(1)} (w)&  =\dfrac{w\bigg(3 r w+4 w^3-ir^{3/2}+\sqrt{r}(e^{-\sqrt{r}|z|}-1)\left[\sqrt{r}w-i(r+2 w^2)\right] \bigg)}{(w+i\sqrt{r})\big(\frac{r}{4}+w^2\big)}, 
\end{align}
and $\mathcal{K}^{\mathcal{A}}_a(x,t)$ is defined in Eq.~\eqref{u-part}.
As done for the instantaneous case we follow the calculations in \ref{AppendixB} to compute $\mathcal{K}_{\sqrt{r}}^{\mathcal{A}^{(1)}}(x,t)$ and find an approximate expression of $\rho(x,t)$ at large $t$ given by
\begin{align}
\rho(x,t) \approx \rho_{ss}(x)\Theta(t\sqrt{4r }-|z|)+\dfrac{e^{-rt-\frac{z^2}{4t}}}{8\pi\sqrt{D}}K^{\mathcal{A}^{(1)}}_{z}\left(\sqrt{r}-\frac{|z|}{2t},t\right),
\label{r-e-d}
\end{align}
where $K_z^{\mathcal{A}}(b,t)$ is given in Eq.~\eqref{J-b-final-mt}. 
A simple observation of the above equation conforms a similar `cone spreading' type relaxation behavior of the density to the steady state.

{After analysing the different terms in Eq.~\eref{r-e-d}, in this case also, one observes that the total density $\rho(x,t)$ satisfies a large deviation form. 
Interestingly, in this case, the LDF is same as that of the instantaneous resetting given in Eq.~\eqref{LD-ins}.}

\begin{figure}[t]
  \begin{center}
  \includegraphics[scale=0.25]{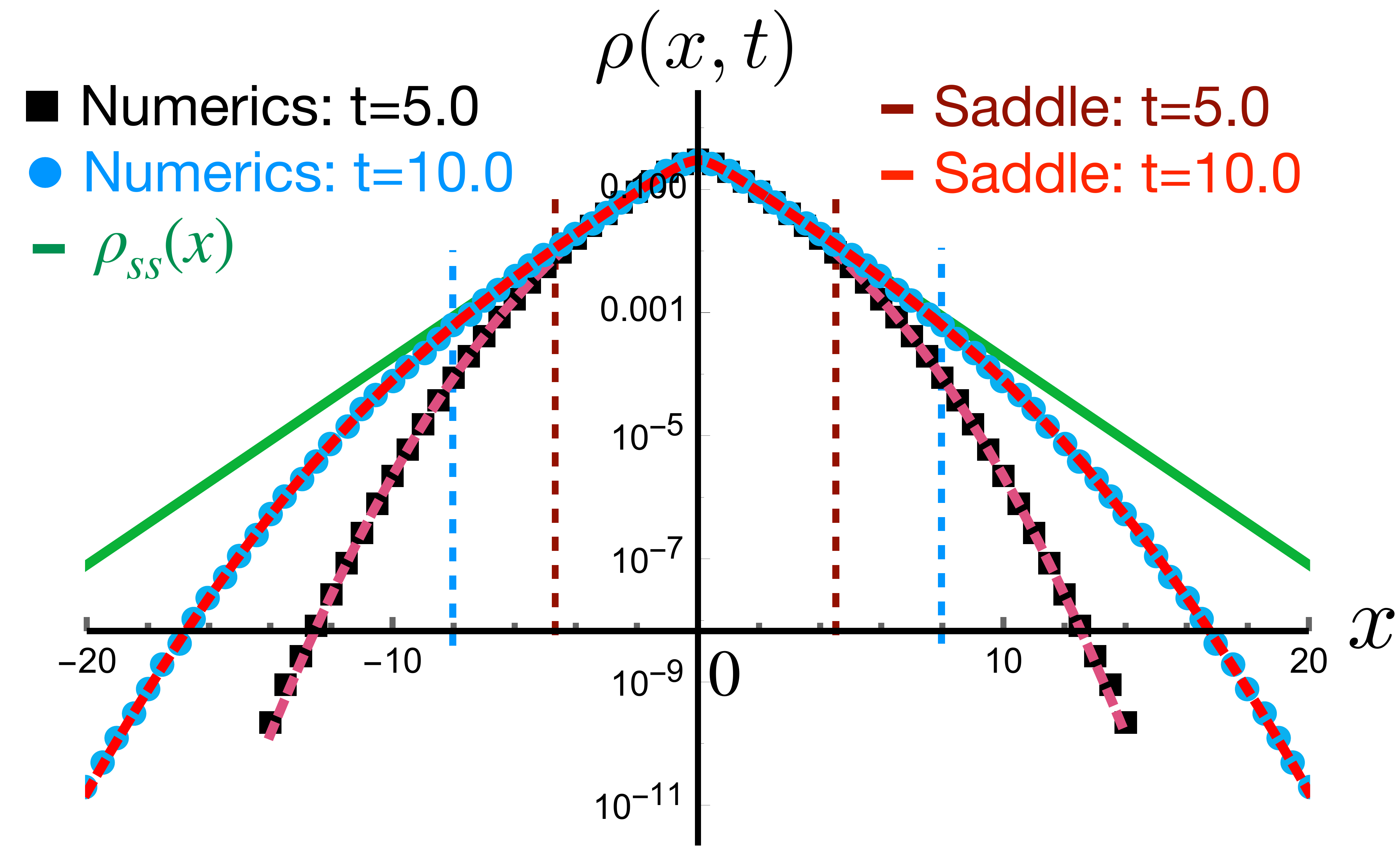}
  \caption{Comparison of the densities $\rho(x,t)$ obtained using saddle point calculation in Eq.~\eqref{relaxation-non-instant-rltbeta} and computing the integrals numerically in Eq.~\eqref{r-lthan-beta} for $r < \beta$. We observe that the distribution within an inner core region (inside the dashed vertical lines given by $x= \pm \sqrt{4Dr}t$) has reached steady state while the outer region is still in the transient state. These two regions are separated by a travelling front which moves ballistically. Parameters used in this plot are: $D=0.5,~r=0.3$ and $\beta=0.8$ i.e., $\lambda=1.2649$. } 
  \label{fig:rho_xt-r-lthan-beta}
  \end{center}
 \end{figure}

\subsubsection{\textbf{Case II ($r<\beta$):~}}
\label{sec:case-2}
{Adding the right hand sides of Eqs.~(\ref{rlbeta-m}) and (\ref{rlbeta-r}), and performing some tedious manipulations we write the total density $\rho(x,t)=\rho_D(x,t)+\rho_R(x,t)$ in the following form}
\begin{align}
\begin{split}
    \rho(x,t) = \rho^{ss}(x) &+
    \frac{i e^{-r t}}{4\pi \sqrt{D}(\sqrt{r+\gamma}-\sqrt{r})}\left[ \mathcal{K}_{\sqrt{r}}^{\mathcal{A}^{(2)}}(x,t)-\mathcal{K}_{\sqrt{r+\gamma}}^{\mathcal{A}^{(2)}}(x,t) \right]
    \\
    &+ \frac{ i re^{-\sqrt{\beta}|z|-\beta t}}{4\pi \lambda^2(\sqrt{\beta+\gamma}-\sqrt{\beta})} \left[ \mathcal{K}_{\sqrt{\beta}}^{\mathcal{A}^{(4)}}(x,t)-\mathcal{K}_{\sqrt{\beta+\gamma}}^{\mathcal{A}^{(4)}}(x,t) \right]
    \\
    &- \frac{r e^{-r t}}{\pi \lambda (\sqrt{r+\gamma}-\sqrt{r})} \text{Im} \left[ i \left(\mathcal{J}_{\sqrt{r}}^{\mathcal{A}^{(3)}}(x,t)-\mathcal{J}_{\sqrt{r+\gamma}}^{\mathcal{A}^{(3)}}(x,t) \right)\right]~,
   \end{split}
   \label{r-lthan-beta} 
\end{align}
where $\gamma \neq 0$ and the $\mathcal{A}^{(i)}(w)$ functions are given in  \ref{A-functions} and $\mathcal{K}^{\mathcal{A}}_a(x,t)$ is defined in Eq.~\eqref{u-part}. In this case we have encountered a new type of integral of the form
\begin{align}
\mathcal{J}^{\mathcal{A}}_a(x,t) = \int_{0}^{\infty}~dw~\frac{\mathcal{A}(w)}{w-ia}~e^{-tw^2+iw|z|},~~~\text{with}~~z=\frac{x}{\sqrt{D}}.  \label{mcalJ^A_a}
\end{align}
This integral is similar to $\mathcal{K}^{\mathcal{A}}_a(x,t)$ except that the integration range is now restricted to $[0, \infty]$. One can follow a similar procedure as done for $\mathcal{K}^{\mathcal{A}}_a(x,t)$ to compute $\mathcal{J}^{\mathcal{A}}_a(x,t)$ for large $t$. In \ref{App-mcalJ} we evaluate this integral using saddle point method and get an approximate expression for large $t$ given by Eq.~\eqref{J_a^A-F}. Evaluating this expression for $\mathcal{A}^{(3)}(w)$ and $\mathcal{K}_{a}^{\mathcal{A}}(x,t)$ from Eq.~\eqref{J_a^A-F} for $\mathcal{A}^{(2)}(w)$ and $\mathcal{A}^{(4)}(w)$, we insert them in Eq.~\eqref{r-lthan-beta} to get an approximate expression for the density $\rho(x,t)$ in the large $t$ limit 
\begin{equation}
    \rho(x,t)\approx~~
    \boxed{
    \begin{aligned}
    &\text{Eq.~\eqref{r-lthan-beta}~with~large~}t~\text{approximations~of~}\mathcal{K}_a^{\mathcal{A}}(x,t)~\text{and}~
    \mathcal{J}_a^{\mathcal{A}}(x,t) \\ 
    &\text{are~inserted~from~Eqs.~\eqref{K_a^A-F}~and~\eqref{J_a^A-F}~respectively.}
   \end{aligned}}
    \label{relaxation-non-instant-rltbeta}
\end{equation}
 From the presence of the $\Theta$ functions in the expressions of $\mathcal{K}_a^{\mathcal{A}}(x,t)$ and 
    $\mathcal{J}_a^{\mathcal{A}}(x,t)$ in Eqs.~\eqref{K_a^A-F}~and~\eqref{J_a^A-F}~respectively, one can see the appearance of many travelling fronts. However, after some manipulations one can see  from the relaxation part of the density in Eq.~\eqref{r-lthan-beta} that  one gets an equal and opposite contributions of the steady state part from terms with $\Theta(|x|-\sqrt{4Dr}t)$, which gets manifested as `cone spreading' relaxation as seen in case of $r =\beta$ in Sec.~\ref{sec:case-1}. Such a relaxation is demonstrated in Fig.~\ref{fig:rho_xt-r-lthan-beta} where we provide a comparison between the saddle point result in Eq.~\eqref{relaxation-non-instant-rltbeta} and the one from direct numerical evaluation of the integrals in Eq.~\eqref{r-lthan-beta}.  We observe nice agreement between them. 
    
{To understand the large deviation behaviour of the density $\rho(x,t)$ in this case, one should look at the large $t$ asymptotic of the integrals of the form $\mathcal{K}_a^{\mathcal{A}}(x,t)$ and $\mathcal{J}_a^{\mathcal{A}}(x,t)$ in Eq.~\eqref{r-lthan-beta}. Before proceeding, we first note that the total contribution coming from these integrals with $a=\sqrt{r+\gamma}$ and $a=\sqrt{\beta+\gamma}$ [{\it i.e.} the total contributions from the second terms in the square brackets in Eq.~\eqref{r-lthan-beta}] is very small because the contributions from the poles at $w= i \sqrt{r +\gamma}$ and $w=i\sqrt{\beta+\gamma}$ should cancel each other. Only the contributions from saddle points survive. This can also be verified by evaluating them in Mathematica. 
This cancellation is expected due to the following observation: recall that these extra poles, in fact, originate from the
pole $s = \gamma$, which appears while performing the inverse Laplace transform of $\tilde{\rho}_D(x,s)$ and $\tilde{\rho}_D(x,s)$ given in  Eqs.~(\ref{rem-1}) and (\ref{rem-2}).  Moreover recall from sec.~\ref{pb-t} 
(see the paragraph after Eq.~\eqref{z(x)}) that $s=\gamma$ is a removable singularity. Hence, the total contributions of these poles at $w = i \sqrt{r +\gamma}$ and $w=i \sqrt{\beta +\gamma}$ from the integrals of the form $\mathcal{K}_a^{\mathcal{A}}(x,t)$ and $\mathcal{J}_a^{\mathcal{A}}(x,t)$ in Eq.~\eqref{r-lthan-beta} should cancel each other and the net sum would be zero. Keeping this in mind and finding the leading contributions of the other three integrals (with poles at $w=i\sqrt{r}$ and $w=i\sqrt{\beta}$) and the saddle point contributions from all the integrals in Eq.~\eqref{r-lthan-beta}, we find that for large $t$, the total density $\rho(x,t)$ satisfies a large deviation form with a LDF which is, interestingly, once again same as that of the instantaneous case given in Eq.~\eqref{LD-ins}. 
Here too, one observes the same dynamical transition similar to the previous cases.
}

\begin{figure}[t]
  \begin{center}
  \includegraphics[scale=0.25]{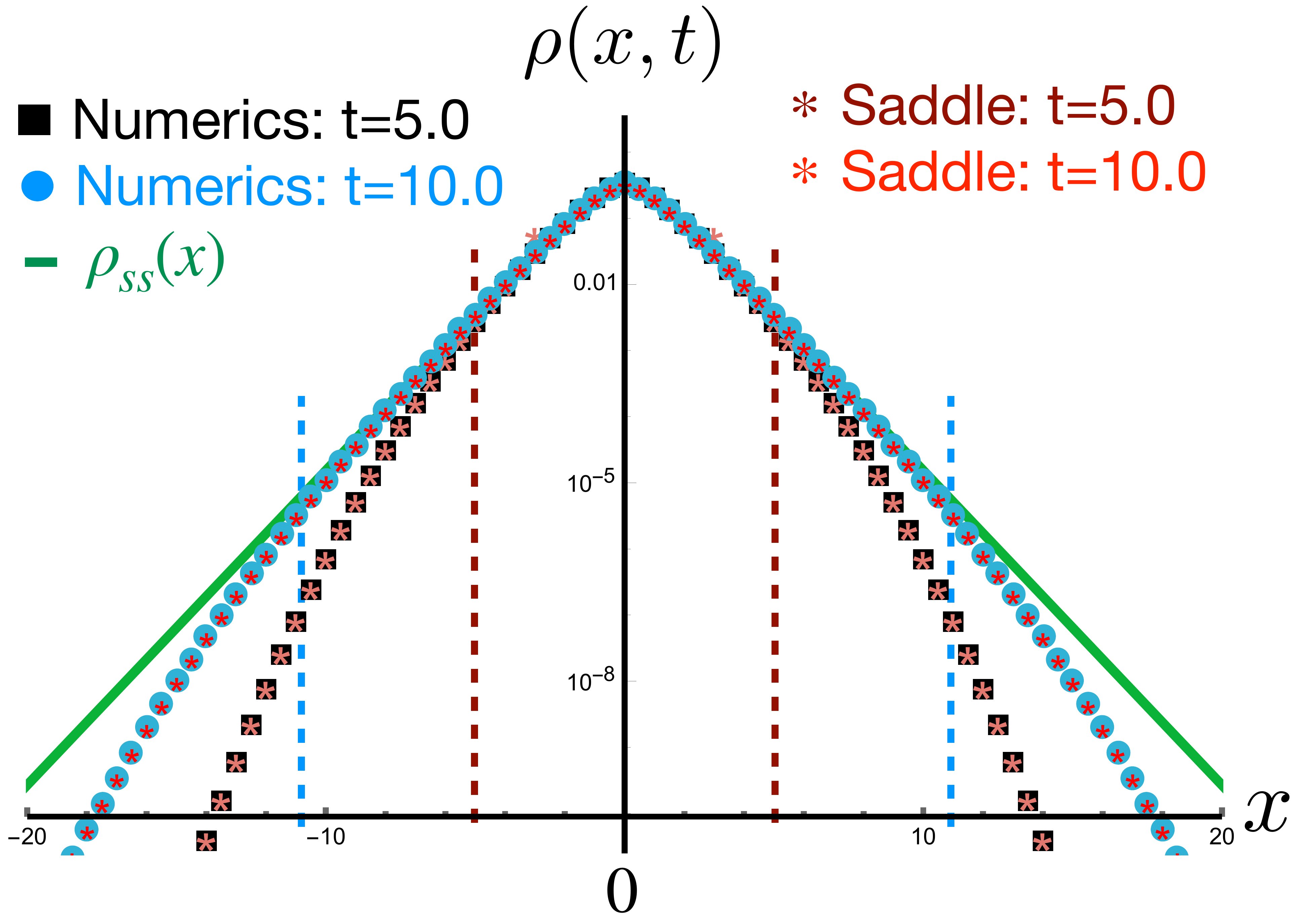}
  \caption{Comparison of the densities $\rho(x,t)$ obtained using saddle point calculation in Eq.~\eqref{relaxation-non-instant-rgtbeta} and computing the integrals numerically in Eq.~\eqref{r-g-beta} for $r > \beta$. We observe that the distribution within an inner core region (inside the dashed vertical lines given by $x= \pm \sqrt{4D\beta}t$)) has reached steady state while the outer region is still in the transient state. This two regions are separated by a travelling front which moves ballistically. Parameters used in this plot are: $D=0.5,~r=0.6$ and $\beta=0.4$ i.e., $\lambda=0.8944$.} 
  \label{fig:rho_xt-r-gthan-beta}
  \end{center}
 \end{figure}

\subsubsection{\textbf{Case II: $r>\beta$}}
{In this case, the total density is obtained by combining Eqs.~\eqref{part-1}-\eqref{part-2} and after some manipulations and rearrangements, we get}
\begin{align}
\begin{split}
\rho(x,t)=\rho^{ss}(x) 
&+\frac{ie^{-r t}}{4 \pi \sqrt{D}(\sqrt{r+\gamma}-\sqrt{r})}\left[ \mathcal{K}^{\mathcal{A}^{(2)}}_{\sqrt{r}}(x,t) - \mathcal{K}^{\mathcal{A}^{(2)}}_{\sqrt{r+\gamma}}(x,t)\right]  \\
&+\frac{ir e^{-\sqrt{\beta} \left| z\right|}e^{-\beta t}}{4 \pi  \lambda ^2(\sqrt{\beta+\gamma}-\sqrt{\beta})} \left[ \mathcal{K}^{\mathcal{A}^{(4)}}_{\sqrt{\beta}}(x,t) - \mathcal{K}^{\mathcal{A}^{(4)}}_{\sqrt{\beta+\gamma}}(x,t)\right] \\
&- \frac{r e^{-r t}}{\pi \lambda (\sqrt{r+\gamma}-\sqrt{r})} \text{Im} \left[ i \left(\mathcal{J}_{\sqrt{r}}^{\mathcal{A}^{(3)}}(x,t)-\mathcal{J}_{\sqrt{r+\gamma}}^{\mathcal{A}^{(3)}}(x,t) \right)\right]  \\
&+\frac{r e^{-r t}}{\pi \lambda(\sqrt{r+\gamma}-\sqrt{r})} \left[ \mathbb{J}_{\sqrt{r},\sqrt{r-\beta}}^{\mathcal{A}^{(3)}}(x,t)- \mathbb{J}_{\sqrt{r+\gamma},\sqrt{r-\beta}}^{\mathcal{A}^{(3)}}(x,t)\right]~, 
\end{split}
\label{r-g-beta} 
\end{align}
{for $\gamma \neq 0$}, where the $\mathcal{A}^{(i)}(w)$ functions are given in  \ref{A-functions}. The functions $\mathcal{K}^{\mathcal{A}}_a(x,t)$ and $\mathcal{J}^{\mathcal{A}}_a(x,t)$ are defined in Eqs.~\eqref{u-part} and \eqref{mcalJ^A_a} respectively. We also define the following integral on real line segment $[0,\zeta]$ for $\zeta>0$ as
\begin{align}
\mathbb{J}_{a,\zeta}^{\mathcal{A}}(x,t)= \int_0^{\zeta} dv~ \frac{\mathcal{A}(iv)}{v-a}e^{tv^2-v|z|},~\text{with}~z=\frac{x}{\sqrt{D}}.
\label{mbbJ_a^A}
\end{align}
This integral naturally appears while evaluating the integral $\mathcal{J}_a^{\mathcal{A}}(x,t)$ in \ref{App-mcalJ}. For $\mathcal{A}(w) = \mathcal{A}^{(3)}(w)$ we analyse this integral in \ref{sec:mbbJ_a^A} where it can be seen using Eq.~\eqref{mbbJ_a^A>} that, a part of the contribution from third line of Eq.~\eqref{r-g-beta} is same to the terms coming from the fourth line but of opposite sign yielding some  cancellations and hence further simplifications [see discussions around  Eqs.~\eqref{mbbJ_a^A>} and \eqref{mbbJ_a^A>-f}]. {Note that the above calculations are done assuming $\gamma \neq0$. One can do a similar calculation for $\gamma=0$ {\it i.e.} $r=4\beta$ and evaluate the integrals using the saddle point method as used in other cases.}

Accumulating all the terms we, for large $t$,  get
\begin{equation}
    \rho(x,t)\approx~~
    \boxed{
    \begin{aligned}
    &\text{Eq.~\eqref{r-g-beta}~with~large~}t~\text{approximations~of~}\mathcal{K}_a^{\mathcal{A}}(x,t),~
    \mathcal{J}_a^{\mathcal{A}}(x,t)~\text{and} \\ 
    &\mathbb{J}_{a,\zeta}^{\mathcal{A}}(x,t)~\text{are~given~in~Eqs.~\eqref{K_a^A-F},~\eqref{J_a^A-F}~and~\eqref{mbbJ_a^A>}~respectively.}
   \end{aligned}}
    \label{relaxation-non-instant-rgtbeta}
\end{equation}
In this case also, many travelling fronts appear due the presence of the $\Theta$ functions in the expressions of $\mathcal{K}_a^{\mathcal{A}}(x,t)$ and $\mathcal{J}_a^{\mathcal{A}}(x,t)$ in Eqs.~\eqref{K_a^A-F}~and~\eqref{J_a^A-F}~respectively. 
Once again, after some manipulations one can see  from the relaxation part of the density in Eq.~\eqref{r-g-beta} that  one gets an equal and opposite contributions of the steady state part from terms with $\Theta(|x|-\sqrt{4D\beta }t)$, which gets manifested as `cone spreading' relaxation as seen in case of $r < \beta$ in Sec.~\ref{sec:case-2}. Such a relaxation is demonstrated in Fig.~\ref{fig:rho_xt-r-gthan-beta} where we provide a comparison between the saddle point result in Eq.~\eqref{relaxation-non-instant-rgtbeta} and the one from direct numerical evaluation of the integrals in Eq.~\eqref{r-g-beta}.  Here too we observe a nice agreement.

\begin{figure}[t]
  \begin{center}
  \includegraphics[scale=0.25]{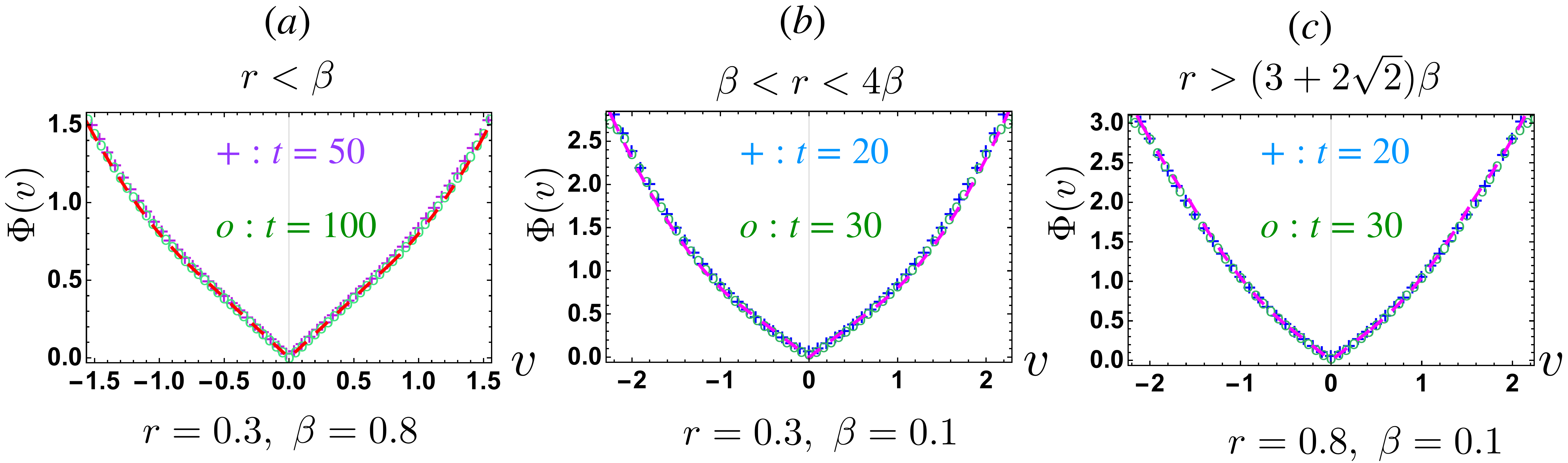}
  \caption{{Plots of the LDF $\Phi(v)$ of the density as given in Eq.~\eqref{LD-all} (dashed lines) for three choices of $r$ and $\beta$. The symbols are obtained from the (saddle point) approximate expressions of the density $\rho(x,t)$ presented in Eqs.~\eqref{relaxation-non-instant-rltbeta} and \eqref{relaxation-non-instant-rgtbeta}.  The diffusion constant used to prepare these plots is $D=0.5$. }
  }
  \label{fig:LDF-all}
  \end{center}
 \end{figure}
 
{Finally we discuss the large deviation behaviour of $\rho(x,t)$ in this case. To get the LDF in this case we follow the same procedure as done in the previous section \ref{sec:case-2}. We find that, unlike the previous cases, in this case the LDF can be different from the instantaneous case. In particular, for $\beta<r<4\beta$ we find the same LDF as in Eq.~\eqref{LD-ins}. But for $r \ge 4 \beta$ we get a different LDF.}

{
Below we summarize the explicit forms of LDF for all values of $r$ and $\beta$: 
\begin{align}
\Phi\left(v\right)&= 
\begin{cases}
&
\begin{sqcases}
\sqrt{\frac{r}{D}} |v|,~&~~~~~\text{for}~|v|< v^* \\
r+\frac{v^2}{4D},~&~~~~~\text{for}~|v|\geq v^* ,
\end{sqcases}
~~~~~~~~~~~~~\text{for}~0<r\leq4\beta, \\
& \\
&
\begin{sqcases}
2\sqrt{\frac{\beta}{D}} |v|,~&\text{for}~|v|< v_1^* \\
\frac{(2\sqrt{\beta D}+|v|)^2}{4D},~&\text{for}~v_1^*\le |v| < v_2^* \\
\sqrt{\frac{r}{D}} |v|,~&\text{for}~v_2^*\le|v|< v^* \\
r+\frac{v^2}{4D},~&\text{for}~|v|\geq v^*
\end{sqcases}
~~~~~~~~\text{for}~4\beta \le r \le (3+2\sqrt{2})\beta, \\
&\\
&
\begin{sqcases}
2\sqrt{\frac{\beta}{D}} |v|,~&\text{for}~|v|< v_1^* \\
\frac{(2\sqrt{\beta D}+|v|)^2}{4D},~&\text{for}~v_1^*\le |v| < v_3^* \\
r+\frac{v^2}{4D},~&\text{for}~|v|\geq v_3^*
\end{sqcases}
~~~~~~~~\text{for}~r \ge (3+2\sqrt{2})\beta>0.\\
\end{cases}
\label{LD-all}
\end{align}
where $v^*=\sqrt{4 D r}$, $v_1^*=\sqrt{4 D \beta}$, $v_2^*=\sqrt{4D}(\sqrt{r}-\sqrt{\beta}+\sqrt{r-\sqrt{4r \beta}})$ and $v_3^*=\sqrt{D}(r-\beta)/\sqrt{\beta}$. Note that for $r=4 \beta$ the points $v_1^*$ and $v_2^*$ coincide. On the other hand,  for $r=(3+2\sqrt{2})\beta$, the three points $v^*,~v_2^*$ and $v_3^*$ coincide.  The above expressions for the LDF in different ranges are plotted in Fig.~\ref{fig:LDF-all}.
Similar to the instantaneous case, the LDF associated to the density $\rho(x,t)$ for finite $\lambda$ also exhibits 
dynamical phase transition at a fixed spatial point as described in Ref.~\cite{transport1}.
}

\section{Summary}
\label{summ}
In this paper, we have studied motion of a Brownian particle under stochastic resetting with finite time return. The return dynamics is facilitated by a linear trap whose minimum is located at the resetting position. The return time is stochastic and it also depends on the position of the particle at the time of resetting. Realizing the fact that motion of the particle can be decomposed into two phases namely the exploration and return phases, we construct Fokker-Planck equations for each phase with suitable boundary conditions. We go beyond the steady state limit, and compute exact time dependent forms of the position density. This also allows us to study the temporal relaxation of the densities at large time, and we show that depending on the relative strength between the resetting rate and potential, different variants of the relaxation forms are found. We have shown that the density has travelling boundaries which separate two regions: an inner core which has already relaxed to the steady state, and an outer core which is still at transient. An important thing to note that the speed at which the boundary between the cores moves depends on the ratio of  the resetting rate and strength of the potential. 

It is worth stressing that in this paper we have taken the master equation approach to discuss the relaxation properties which, to the best of our knowledge, was only shown before using the renewal formalism for the instantaneous resetting. While the latter approach in the current set-up is still a far-off, the master equation formalism allows us to make a connection between the singularities in the densities in Laplace space and their relaxation properties. We benchmark this approach first for the instantaneous case, and then employ it in the case of non-instantaneous stochastic returns.

Computation of time dependent distributions for a stochastic process in the presence of resetting is rather limited unlike that of the non-equilibrium steady state. In this paper, we have provided a systematic way to compute these probability densities. Various physical limits can be reached from our formulation -- thus, we believe that the methods shown here can be very useful to compute propagators and study relaxation in other resetting systems. Finally, we mention that one interesting future direction would be to study the effect of non-instantaneous stochastic return on the first passage problems.

\section{Acknowledgement}
We thank Carlos A. Plata for many fruitful discussions.
DG acknowledges the support from the University of
Padova through ``Excellence Project 2018'' of the Fondazione
Cassa di Risparmio di Padova e Rovigo. AP gratefully acknowledges
Raymond and Beverly Sackler postdoctoral fellowship and Ratner Center for Single
Molecule Science at the Tel Aviv University for funding.
 AK would like to acknowledge support of the Department of Atomic Energy, Government of India, under project no.12-R\&D-TFR-5.10-1100 and the support from DST, Government of India grant under project No. ECR/2017/000634. 

\appendix
\section{Inverse Laplace transform of Eqs. \eqref{rem-1} and \eqref{rem-2}}
\label{sec:bc-int}
\begin{figure}[t]
  \begin{center}
    \includegraphics[width=8cm]{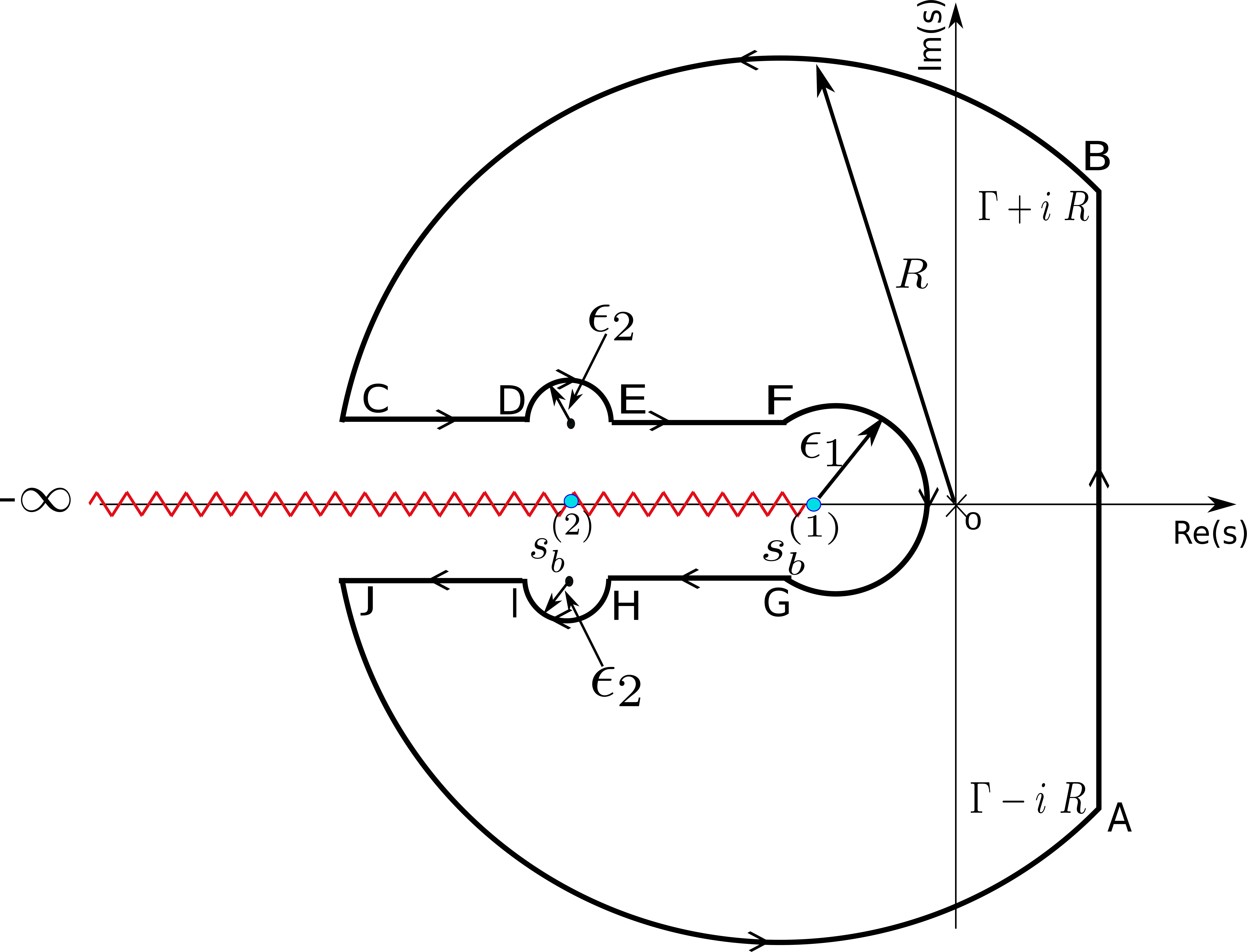}
    \caption{Bromwich contour for inverse Laplace transform in the complex $s-$plane. Two branch point singularities at $s_b^{(1)}$ and $s_b^{(2)}<s_b^{(1)}<0$ and a pole at $s=0$. The pole is indicated by cross and branch points are shown by filled circles. Zigzag curves represent the branch cuts that connect the branch points to the branch point at $s=-\infty$.  }
    \label{fig:BC}
  \end{center}
\end{figure}
\noindent
In this section, we provide details of inverse Laplace transforms of  Eqs. (\ref{rem-1}) and (\ref{rem-2}) to get the densities in the exploration and return phases as given in Eqs.~(\ref{rlbeta-m}-\ref{reqbeta-r}). We start by recalling $\tilde{\rho}_{D,R}(x,s)$ below as 
\begin{align}
\tilde\rho_D(x,s)&=\dfrac{\lambda(r+2 s)+r\sqrt{4D} \sqrt{s+\beta}}{4 D s(\lambda \sqrt{(s+r)/D}+r)}~e^{-\sqrt{s+r}|z|},\label{rem-1-a}\\
\tilde\rho_R(x,s)&=r\dfrac{\lambda(r+2 s)+ r\sqrt{4D} \sqrt{s+\beta}}{4 \lambda^2 s(s-\gamma)}~\big(e^{-\sqrt{s+r}|z|}-e^{-\sqrt{\beta} |z|}~e^{-\sqrt{s+\beta}|z|}\big)\label{rem-2-a},
\end{align}
where  $z=\frac{x}{\sqrt{D}}$, $\beta=\frac{\lambda^2}{4 D}$, and $\gamma=-r+\frac{r^2}{4\beta}$. 
Following the Bromwich  integral formula and the residue theorem, the inverse Laplace transform of $\tilde{\rho}_{D,R}(x,s)$ can be
formally written as
\begin{align}
\rho_{D,R}(x,t)=\dfrac{1}{2 \pi i}\int_{\Gamma-i\infty}^{\Gamma+i\infty}~ds~e^{st}~\tilde{\rho}_{D,R}(x,s)\label{ILT},
\end{align}
where the vertical contour passes through $\text{Re}(s)=\Gamma$ in the complex $s$-plane such that all the singularities lie to left of it (see e.g., Fig.~\ref{fig:BC}). Both the functions $\tilde{\rho}_{D,R}(x,s)$ have a simple pole at $s=0$ and three branch points: $s=-r, s=-\beta$ and $s=-\infty$. 



To compute the Bromwich integral in \eref{ILT} in this case, we consider the contour integral  
$\dfrac{1}{2\pi i}\oint_{\mathcal{C}}~ds~e^{s t}~\tilde{\rho}_{D,R}(x,s) $
along the contour $\mathcal{C}$ shown in Fig.~\ref{fig:BC}. It is straightforward to see that it  has the value 
\begin{align}
\dfrac{1}{2\pi i}\oint_{\mathcal{C}}~ds~e^{s t}~\tilde{\rho}_{D,R}(x,s) 
&=\text{Residue}[s=0],
\label{std-int}
\end{align}
 The integral along the contour $\mathcal{C}$ in  Fig.~\ref{fig:BC} can be decomposed in the following manner 
\begin{align}
\begin{split}
\dfrac{1}{2 \pi i} \int_{\mathcal{C}}~ds~e^{s t}~\tilde{\rho}_{D,R}(x,s) &= \\
&=\dfrac{1}{2 \pi i} \left[ \int_{AB}+\int_{BC}+\int_{CD}+\int_{DE}+\int_{EF} \right.  \\ 
&~~~~~ \left. +\int_{FG} +\int_{GH}+\int_{HI}+\int_{IJ}+\int_{JA} \right] ds~e^{st}~\tilde{\rho}_{D,R}(x,s).  \\
&=\text{Residue}[s=0]
\end{split}
\label{brom}
\end{align}
Note that the inverse Laplace transform in Eq.~\eqref{ILT} is actually the integral over the part `AB' on the contour.
In the limits of $R\to\infty$ and $\epsilon_{1,2}\to 0$, one can show that the integrals along the
chords `BC', `DE', `FG', `HI' and `JA' vanish. Hence,  we get 
\begin{align}
\rho_{D,R}(x,t)=\frac{1}{2\pi i} \int_AB = \text{Residue}[s=0] - \frac{1}{2 \pi i} \left[ \int_{CD} +\int_{EF} +\int_{GH} +\int_{IJ}  \right]\label{ILT-1},
\end{align}
where the integrals along `CD', and `EF' are performed on the negative real axis above the branch cut whereas the integrals along `GH', and `IJ' are performed below the branch cut. 
Writing these integrals in terms of real integration variables and performing some manipulations, {we get the following expressions} for $\rho_D(x,t)$ and $\rho_R(x,t)$
 for different choices of $r$ and $\lambda$ (equivalently $\beta$).

{
\subsection{$r<\beta$} 
For the density $\rho_{D}(x,t) $ in the exploration phase, we get  
\begin{align}
\begin{split}
&\rho_{D}(x,t) 
=\rho^{ss}_{D}(x)+\frac{1}{4 \pi  \lambda ^2 \sqrt{D}}\int_0^{\beta-r}~dy~\frac{e^{-t(r +y)}}{(r +y) (r +\gamma+y)} \\
&~
\times \left[r \left(\lambda -2 \sqrt{D(\kappa-y)}\right)+2 \lambda  y\right] \left[\lambda  \sqrt{y} \cos \left(\sqrt{y} \left| z\right| \right)+r\sqrt{D} \sin \left(\sqrt{y} \left| z\right| \right)\right]\\
&~+\frac{1}{4 \pi  \sqrt{D} \lambda ^2}\int_0^\infty~dy~\frac{e^{-t(\beta +y)}}{(\beta +y) (\beta +\gamma+y)}\bigg[\left\{2 D r^2 \sqrt{y}+\lambda ^2 (2 (\beta +y)-r) \sqrt{\kappa+y}\right\}  \\
&~\times\cos \left(\left| z\right|  \sqrt{\kappa+y}\right)-r\lambda\sqrt{D} \left\{2 \sqrt{y(\kappa+y)}+r-2 (\beta +y)\right\} \sin \left(\left| z\right|  \sqrt{\kappa+y}\right)\bigg],
\end{split}
\label{rlbeta-m}
\end{align}
 where recall $z=x/\sqrt{D}$ and  $\kappa=\beta-r>0$. 
 {On the other hand for the density in the return phase, we get}
\begin{align}
&\rho_{R}(x,t)
=\rho^{ss}_{R}(x)\nonumber\\
&+\frac{r}{4 \pi  \lambda ^2}\int_0^{\beta-r}~dy~\frac{e^{-(r+y)t}}{(r+y) (r+\gamma+y)}\bigg[\left\{2 r\sqrt{D(\kappa-y)}-\lambda  r-2 \lambda y\right\} \sin \left(\sqrt{y} \left| z\right| \right) \bigg] 
\nonumber \\
&+\frac{r}{4 \pi  \lambda ^2}\int_0^\infty~dy~\frac{e^{-t(\beta +y)} }{(\beta +y) (\beta +\gamma+y)}\bigg[2 r \sqrt{D y} \left\{e^{-\sqrt{\beta}|z|} \cos \left(\sqrt{y} \left| z\right| \right) \right. \label{rlbeta-r} \\
&~-\left.\cos \left(\left| z\right|  \sqrt{\kappa+y}\right)\right\}-\lambda  [r-2 (\beta +y)] \left\{e^{-\sqrt{\beta}|z|} \sin \left(\sqrt{y} \left| z\right| \right)-\sin \left(\left| z\right|  \sqrt{\kappa+y}\right)\right\}\bigg], \nonumber
\end{align}
where $z=x/\sqrt{D}$ and the densities in the steady state are given in Eqs.~\eqref{ss-1} and \eqref{ss-2}.

\subsection{$r>\beta$} 
Similar prescription also  applies for  $r>\beta$. Following the steps described above, we find
\begin{align}
&\rho_{D}(x,t)=\rho^{ss}_{D}(x)
+\frac{1}{2 \pi \lambda^2}\int_0^{\Lambda}~dy~\frac{r \sqrt{Dy}~e^{-t(\beta +y)}e^{-\left| z\right|  \sqrt{\Lambda-y}}}{(\beta +y) (\beta +\gamma+y)}\left(r-\lambda  \sqrt{\frac{\Lambda-y}{D}}\right)
\nonumber\\
&~~~~+\frac{1}{4 \pi \lambda^2}\int_0^\infty~dy~\frac{e^{-t(r +y)}}{(r +y) (r +\gamma+y)}\bigg[\cos \left(\sqrt{y} \left| z\right| \right) \left(2 r^2 \sqrt{D (\Lambda+y)} +\lambda ^2 r \sqrt{\frac{y}{D}} \right. \label{part-1}  \\
&~~~~
\left.+~2 \lambda ^2 y \sqrt{\frac{y}{D}}\right) +\lambda  r \sin \left(\sqrt{y} \left| z\right| \right) \{-2 \sqrt{y (\Lambda+y)}+r+2 y\} \bigg], \nonumber \\
&\rho_{R}(x,t)=\rho^{ss}_{R}(x)+
+\frac{r}{4 \pi  \lambda ^2}\int_0^{\Lambda}~dy~\frac{e^{-t(\beta +y)}}{(\beta +y) (\beta +\gamma+y)}\bigg[2 r \sqrt{D y} \left\{e^{-\sqrt{\beta}|z|} \cos \left(\sqrt{y} \left| z\right| \right) \right.
\nonumber\\
&\left.-\cosh \left(\left| z\right|  \sqrt{\Lambda-y}\right)\right\}+e^{-\sqrt{\beta}|z|} \lambda  [2 (\beta +y)-r] \sin \left(\sqrt{y} \left| z\right| \right) +2 r \sqrt{D y} \sinh \left(\left| z\right|  \sqrt{\Lambda-y}\right)\bigg]\nonumber\\
&-\frac{r}{4 \pi  \lambda ^2}\int_0^\infty~dy~\frac{e^{-t(r +y)}}{(r +y) (r +\gamma+y)}\bigg[2 r \sqrt{D(\Lambda+y)} \left\{\cos \left(\sqrt{y} \left| z\right| \right) \right. \label{part-2}\\
&\left.-e^{-\sqrt{\beta}|z|} \cos \left(\left| z\right|  \sqrt{\Lambda+y}\right)\right\}+\lambda  (r+2 y) \left\{\sin \left(\sqrt{y} \left| z\right| \right)-e^{-\sqrt{\beta}|z|} \sin \left(\left| z\right|  \sqrt{\Lambda+y}\right)\right\}\bigg],
\nonumber
\end{align}
where $z=x/\sqrt{D}$ and $\Lambda=-\kappa=r-\beta>0$. 

}

\section{Evaluation of the integral $\mathcal{K}^{\mathcal{A}}_a(x,t)$ appearing  in Eq. \eqref{u-part}} 
\label{AppendixB}
\begin{figure}[t]
    \centering
    \includegraphics[width=10cm]{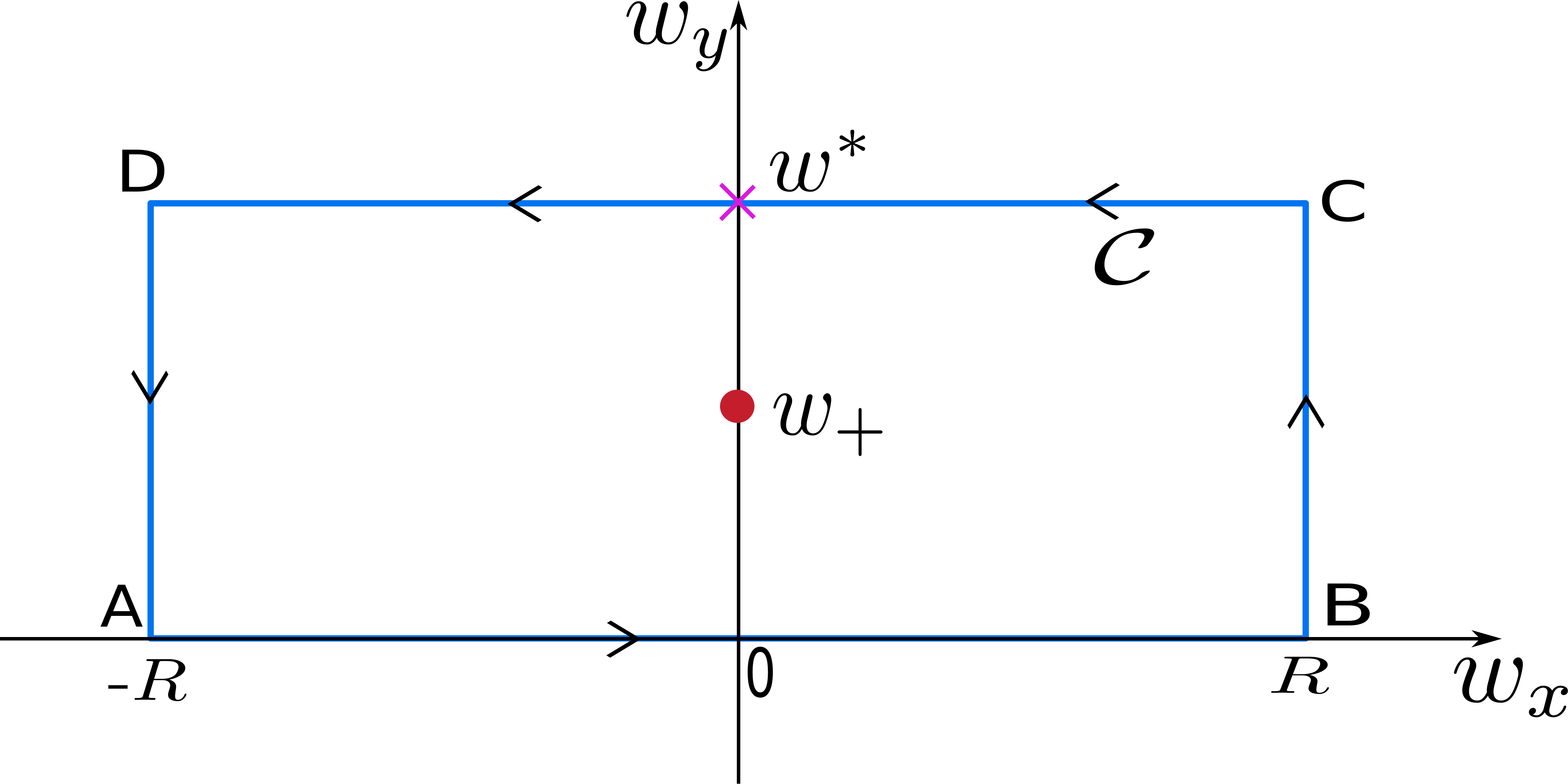}
    \caption{Closed contour $\mathcal{C}$ for the integral \eqref{int-3}. $w^*$ and $w_+$, respectively, are the saddle point and pole, and both of them are function of a variable $z$. The horizontal path through the saddle point $w^*$ is the steepest decent path.}
    \label{fig:contour-C}
\end{figure}
In this section, we evaluate the integral of the form 
\begin{align}
    \mathcal{K}_{a}^{\mathcal{A}}(x,t) &=  \int_{-\infty}^{+\infty}~dw~\frac{\mathcal{A}(w)}{w-ia}~e^{-tw^2+iw|z|},~~a>0, \label{int-28}
\end{align}
where $z=x/\sqrt{D}$. To compute the above integral, let us consider the following integral over a closed contour $\mathcal{C}$ as shown in Fig.~\ref{fig:contour-C} in the complex plane:
\begin{align}
\begin{split}
\oint_\mathcal{C} ~dw~\frac{\mathcal{A}(w)}{w-ia}~e^{-tw^2+iw|z|} &= 2 \pi i~\text{Residue}[ia]~\Theta(|z| - 2 a t) \\ 
&= 2 \pi i~\mathcal{A}(ia)~e^{a^2 t-a|z|} \Theta(|z|-2at),
\end{split}
\label{int-3}
\end{align}
where we have assumed that $\mathcal{A}(w)$ does not have any singularity inside the contour $\mathcal{C}$. We choose this particular contour because, as can be easily seen that, the horizontal line parallel to the real axis passes
through the saddle point $w^*=\frac{i|z|}{2t}$ and is actually the steepest-descent path. It is easy to identify that the desired integral $\mathcal{K}_{a}^{\mathcal{A}}(x,t) $ in Eq.~\eqref{int-28} corresponds to the integral $\int_{AB}$. Since, in the $R \to \infty$, the integrals $\int_{BC}$ and $\int_{DA}$ go to zero, we get $ \mathcal{K}_{a}^{\mathcal{A}}(x,t) = \int_{AB} = 2 \pi i~\text{Residue}[ia]~\Theta(|z| - 2 a t)-\int_{CD}$. Hence
\begin{align}
 \mathcal{K}_{a}^{\mathcal{A}}(x,t) &= 2 \pi i~\mathcal{A}(ia)~e^{a^2 t-a|z|}~\Theta(|z| - 2 a t)+e^{-\frac{z^2}{4t}} ~K_z^{\mathcal{A}}\left(a-\frac{|z|}{2t},t\right),\label{K_a^A} \\
 \text{where} & ~
 K_z^{\mathcal{A}}(b,t) = \int_{-\infty}^\infty~dw~\frac{\mathcal{A}\left(w+i{|z|}/{2t}\right)}{w-ib}~e^{-tw^2}.
    \label{J-b-def}
\end{align}
The integral $ K_z^{\mathcal{A}}(b,t)$ in the above equation can be evaluated for large $t$ using saddle point method \cite{wong,Sanjib}. However, we need to be careful when the saddle point $w^*=0$ is close to the pole at $w_+=ib$. To proceed, one needs to separate the singular and non-singular part of the integrand. For this we rewrite  
\begin{align}
 K_z^{\mathcal{A}}(b,t) &=\mathcal{A}\left(ib+i\frac{|z|}{2t}\right) \int_{-\infty}^\infty~dw~\frac{e^{-tw^2}}{w-ib}+\int_{-\infty}^\infty~dw~\psi(w)~e^{-tw^2} , \label{J-b} \\ 
\text{where}~&~\psi(w) = \frac{\mathcal{A}\left(w+i {|z|}/{2t}\right)-\mathcal{A}\left(ib+i {|z|}/{2t}\right)}{w-ib}. \label{psi-w}
\end{align}
Note that now the function $\psi(w)$ does not have any singularity since $\mathcal{A}(w)$ is an analytic function on the upper half plane. Following \cite{wong,Sanjib}, we evaluate the first integral by defining
\bea
L_f(b,t)&=\int_{-\infty}^\infty~dw~\frac{e^{-tw^2}}{w-ib},~~\text{such~that,}
\label{J-1} 
\\
K_z^{\mathcal{A}}(b,t) &= \mathcal{A}\left(ib+i\frac{|z|}{2t}\right) ~L_f(b,t) +\int_{-\infty}^\infty~dw~\psi(w)~e^{-tw^2}. \label{J^A-L_f}
\eea
Note that the function $L_f(b,t)$ satisfies the following differential equation 
\bea
\frac{dL_f}{dt}= b^2L_f(t) - i b \sqrt{\frac{\pi}{t}},
\eea
which can be solved with boundary condition  $L_f(\infty)=0$. We get,
\bea
L_f(b,t)&=ib e^{tb^2}~\int_t^\infty~d\tau~e^{-\tau b^2}~\sqrt{\frac{\pi}{\tau}}  =i\pi e^{tb^2} \text{sgn}(b) ~\text{erfc}(|b|\sqrt{t}).
\eea
On the other hand the  integral in Eq.~\eqref{J^A-L_f} for large $t$ can be evaluated straightforwardly using Laplace method and one gets a contribution $\psi(0)\sqrt{\frac{\pi}{t}}$ in the leading order. The next order terms can be found easily by expanding $\psi(w)$ in a Taylor series around $w=0$.  Collecting both the contributions, we finally get an explicit expression of $ K_z^{\mathcal{A}}(b,t)$ at large $t$.  Inserting this expression  in Eq.~\eqref{K_a^A} one gets an  explicit expression of $\mathcal{K}_{a}^{\mathcal{A}}(x,t)$ as 
\begin{align}
\begin{split}
 \mathcal{K}_{a}^{\mathcal{A}}(x,t) &= 2 \pi i~\mathcal{A}(ia)~e^{a^2 t-a|z|}~\Theta(|z| - 2 a t)+e^{-\frac{z^2}{4t}} ~K_z^{\mathcal{A}}\left(a-\frac{|z|}{2t},t\right), \\
 \text{where,}  ~
 K_z^{\mathcal{A}}(b,t) &\approx \mathcal{A}\left(ib+i\frac{|z|}{2t}\right)e^{tb^2} \left[ i\pi~\text{sgn}(b) ~\text{erfc}(|b|\sqrt{t})\right]+\psi(0)\sqrt{\frac{\pi}{t}}~,
 \end{split}
\label{K_a^A-F}
\end{align}
with $\psi(0)=\frac{i}{b}\left[ \mathcal{A}\left(i{|z|}/{2t}\right)-\mathcal{A}\left(ib+i{|z|}/{2t}\right)  \right]$.

\section{Explicit forms of $\mathcal{A}^{(i)}(w)$ functions}
\label{A-functions}
Here, we present the explicit forms for the $\mathcal{A}^{(i)}$ functions which were introduced in the main text
\begin{align}
   \mathcal{A}^{(2)} (w)&=\frac{w^2(2w^2+r)}{(w+i\sqrt{r})(w+i\sqrt{r+\gamma})}, \label{A^(2)}\\
      \mathcal{A}^{(3)} (w)&=\frac{w^2\sqrt{w^2+r-\beta}}{(w+i\sqrt{r})(w+i\sqrt{r+\gamma})}, ~\label{A^(3)} \\
   \mathcal{A}^{(4)} (w)&=\frac{\sqrt{4r^2D}w^2-i\lambda w(2\beta+2w^2-r)}{(w+i\sqrt{\beta})(w+i\sqrt{\beta+\gamma})}. ~\label{A^(4)}
\end{align}

\begin{figure}[t]
    \centering
    \includegraphics[width=10cm]{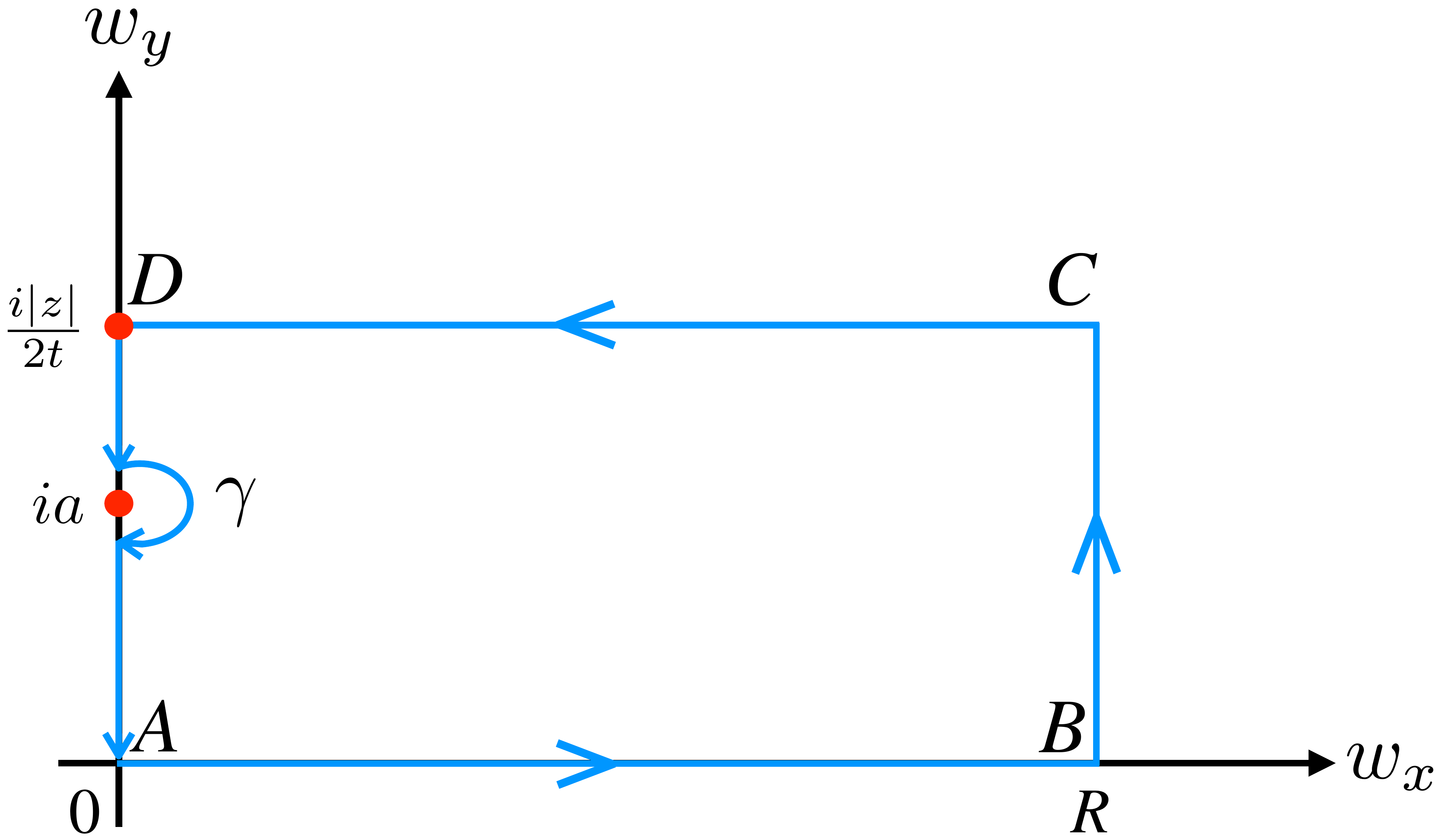}
    \caption{Closed contour $\mathcal{C}$ for the integral \eref{mcalJ^A_a-A}. The saddle point is located at $\frac{i|z|}{2t}$, while the semi-circular integral contributes for $|z| \ge 2at$.}
    \label{fig:contour-C-half}
\end{figure}

\section{Evaluation of the integral $\mathcal{J}^{\mathcal{A}}_a(x,t) $ appearing in Eq. \eqref{mcalJ^A_a}} 
\label{App-mcalJ}
In this section, we evaluate the integral of the form 
\begin{align}
\mathcal{J}^{\mathcal{A}}_a(x,t) = \int_{0}^{\infty}~dw~\frac{\mathcal{A}(w)}{w-ia}~e^{-tw^2+iw|z|},~~~\text{with}~~z=\frac{x}{\sqrt{D}}, \label{mcalJ^A_a-A}
\end{align}
and $a>0$. To compute the above integral, we consider the following integral over a closed contour $\mathcal{C}$ as shown in Fig.~\ref{fig:contour-C-half} in the complex plane:
\begin{align}
\oint_\mathcal{C} ~dw~\frac{\mathcal{A}(w)}{w-ia}~e^{-tw^2+iw|z|} = 0,\label{int-mcalJ}
\end{align}
where we have assumed that $\mathcal{A}(w)$ does not have any singularity inside the contour $\mathcal{C}$. It is easy to identify that the desired integral $\mathcal{J}_{a}^{\mathcal{A}}(x,t) $ in Eq.~\eqref{mcalJ^A_a-A} corresponds to the integral $\int_{AB}$. Since, in the $R \to \infty$, the integral $\int_{BC}$  goes to zero, we get 
\begin{align} 
\mathcal{J}_{a}^{\mathcal{A}}(x,t) &= \int_{AB} = -\int_{\gamma}-\int_{DA}-\int_{CD} 
= \frac{1}{2}\ointctrclockwise_{w=ia} + \int_{AD}+\int_{DC}, \label{mcalJ_a^A}
\end{align} 
where the integral $\int_{AD}$ represents integral along $AD$. 
The integrals on the small circle around $w=ia$ can be evaluated easily and one gets 
\begin{align}
 \frac{1}{2}\ointctrclockwise_{w=ia} = i \pi \mathcal{A}(ia) e^{a^2t-a|z|}~\Theta(|z|-2at).
\end{align}
On the other hand, the integral $\int_{AD} $ on AD is exactly the integral $\mathbb{J}_{a,\zeta}^{\mathcal{A}}(x,t)$ defined in Eq.~\eqref{mbbJ_a^A} with $\zeta=|z|/2t$ and is written explicitly as 
\begin{align}
\mathbb{J}_{a, \frac{|z|}{2t}}^{\mathcal{A}}\left(x,t\right)= \int_0^{\frac{|z|}{2t}} dv~ \frac{\mathcal{A}(iv)}{v-a}e^{tv^2-v|z|}.\label{int_AD}
\end{align}
Note that this is an integral over $v$ within the finite range $[0,|z|/2t]$ on the real line.
Collecting all the contributions in Eq.~\eqref{mcalJ_a^A} explicitly, we get
\begin{align}
\begin{split}
 \mathcal{J}_{a}^{\mathcal{A}}(x,t) &= i \pi \mathcal{A}(ia) e^{a^2t-a|z|}~\Theta(|z|-2at)  + \mathbb{J}_{a, \frac{|z|}{2t}}^{\mathcal{A}}\left(x,t\right)+e^{-\frac{z^2}{4t}} ~J_z^{\mathcal{A}}\left(a-\frac{|z|}{2t},t\right),
\end{split}
\label{J_a^A} \\
 \text{where,} & ~
 J_z^{\mathcal{A}}(b,t) = \int_{0}^\infty~dw~\frac{\mathcal{A}\left(w+i\frac{|z|}{2t}\right)}{w-ib}~e^{-tw^2}.
    \label{J-b-def-half}
\end{align}
We now proceed to evaluate the integral $J_z^{\mathcal{A}}(b,t)$ in the above equation for large $t$ using saddle point method in  similar way as done in \ref{AppendixB}. Once again we need to be careful when the saddle point $w^*=0$ is close to the pole at $w_+=ib$. To proceed, one needs to separate the singular and non-singular part of the integrand. For this we rewrite  
\begin{align}
 J_z^{\mathcal{A}}(b,t) &=\mathcal{A}\left(ib+i\frac{|z|}{2t}\right) \int_{0}^\infty~dw~\frac{e^{-tw^2}}{w-ib}+\int_{0}^\infty~dw~\psi(w)~e^{-tw^2} , \label{J-b} \\ 
\text{where,}~~\psi(w) &=  \frac{\mathcal{A}\left(w+i {|z|}/{2t}\right)-\mathcal{A}\left(ib+i {|z|}/{2t}\right)}{w-ib}.\label{psi-w-half}
\end{align}
We evaluate the first integral by defining
\bea
L_h(b,t)&=\int_{0}^\infty~dw~\frac{e^{-tw^2}}{w-ib},~~\text{such~that,}
\label{J-1} \\
J_z^{\mathcal{A}}(b,t) &= \mathcal{A}\left(ib+i\frac{|z|}{2t}\right) L_h(b,t) +\int_{0}^\infty~dw~\psi(w)~e^{-tw^2}  . \label{J^A-L_f-II}
\eea
Note that the function $L_h(b,t)$ satisfies the following differential equation 
\bea
\begin{split}
\frac{dL_h}{dt}&=b^2L_h(t)-\int_{0}^\infty~dw~(w+ib)~e^{-tw^2} \\
&= b^2L_h(t)- \left( \frac{1}{2 t}+\frac{ib}{2}\sqrt{\frac{\pi}{t}}  \right),
\end{split}
\eea
which can be solved with boundary condition  $L_f(\infty)=0$. We get,
\bea
L_h(b,t)&=e^{tb^2}~\int_t^\infty~d\tau~e^{-\tau b^2}~\left(\frac{1}{2\tau}+\frac{ib}{2}\sqrt{\frac{\pi}{\tau}}    \right) \nonumber\\
&=\frac{e^{tb^2}}{2}\left[ \Gamma(0,tb^2)+i\pi \text{sgn}(b) ~\text{erfc}(|b|\sqrt{t})  \right],
\eea
where $\Gamma(0,z)=\int_z^\infty~du~\frac{e^{-u}}{u}$ is an incomplete Gamma function.
On the other hand the second integral in Eq.~\eqref{J^A-L_f-II} for large $t$ can be evaluated straightforwardly using Laplace method 
because  the function $\psi(w)$ in Eq.~\eqref{psi-w-half} does not have any singularity for $\mathcal{A}(w)$ being analytic function on the upper half plane. Hence from the second integral one gets a contribution $\frac{\psi(0)}{2}\sqrt{\frac{\pi}{t}}$ at large $t$.  
Now, collecting both the contributions, we finally get an explicit expression of $ J_z^{\mathcal{A}}(b,t)$ at large $t$.  Inserting this expression  in Eq.~\eqref{J_a^A} one gets an  explicit expression of $\mathcal{J}_{a}^{\mathcal{A}}(x,t)$ as 
\begin{align}
\begin{split}
 \mathcal{J}_{a}^{\mathcal{A}}(x,t) &= i \pi \mathcal{A}(ia) e^{a^2t-a|z|}~\Theta(|z|-2at)  + \mathbb{J}_{a, \frac{|z|}{2t}}^{\mathcal{A}}\left(x,t\right)
+e^{-\frac{z^2}{4t}} ~J_z^{\mathcal{A}}\left(a-\frac{|z|}{2t},t\right),\\
\text{where,} &  \\
 J_z^{\mathcal{A}}(b,t) &\approx \mathcal{A}\left(ib+i\frac{|z|}{2t}\right)\frac{e^{tb^2}}{2} \left[ \Gamma(0,tb^2)+ i\pi~\text{sgn}(b) ~\text{erfc}(|b|\sqrt{t})\right]+\frac{\psi(0)}{2}\sqrt{\frac{\pi}{t}}~,
 \end{split}
\label{J_a^A-F}
\end{align}
with $\psi(0))=\frac{i}{b}\left[ \mathcal{A}\left(i{|z|}/{2t}\right)-\mathcal{A}\left(ib+i{|z|}/{2t}\right)  \right]$ and $\mathbb{J}_{a, \frac{|z|}{2t}}^{\mathcal{A}}(x,t)$ is defined in Eq.~\eqref{int_AD}. 

\section{Remarks on the integral $\mathbb{J}_{a, \frac{|z|}{2t}}^{\mathcal{A}^{(3)}}\left(x,t\right)$ defined in Eq.~\eqref{int_AD}}
\label{sec:mbbJ_a^A}
Since in Eqs.~\eqref{r-lthan-beta} and \eqref{r-g-beta}, we require to evaluate $\mathcal{J}_{a}^{\mathcal{A}}(x,t)$ for $\mathcal{A}(w)=\mathcal{A}^{(3)}(w)$, we here rewrite the function $\mathbb{J}_{a, \frac{|z|}{2t}}^{\mathcal{A}}(x,t)$ [given in Eq.~\eqref{int_AD}] with $\mathcal{A}=\mathcal{A}^{(3)}(w)$ for $r<\beta$ and $r>\beta$.
 For $r < \beta$, the integral $\mathbb{J}_{a,\frac{|z|}{2t}}^{\mathcal{A}^{(3)}}(x,t)$ reads
\begin{align}
\mathbb{J}_{a, \frac{|z|}{2t}}^{\mathcal{A}^{(3)}}\left(x,t\right)= 
 &~~  i \int_{0}^{\frac{|z|}{2t}} dv~ \frac{v^2\sqrt{v^2+\beta-r}}{(v-a)(v+\sqrt{r})(v+\sqrt{r+\gamma})}e^{tv^2-v|z|}, \label{mbbJ_a^A<}
\end{align}
where we have used the expression of $\mathcal{A}^{(3)}(w)$ in Eq.~\eqref{A^(3)}. Note that in this case $\mathbb{J}_{a, \frac{|z|}{2t}}^{\mathcal{A}}(x,t)$ is purely imaginary and hence will not finally contribute in Eq.~\eqref{r-lthan-beta}. 

On the other hand for $r>\beta$, we rewrite the integral as 
\begin{align}
\begin{split}
\mathbb{J}_{a, \frac{|z|}{2t}}^{\mathcal{A}^{(3)}}\left(x,t\right)= 
 &~~  i \int_{\min(\sqrt{r-\beta},\frac{|z|}{2t})}^{\frac{|z|}{2t}} dv~ \frac{v^2\sqrt{v^2+\beta-r}}{(v-a)(v+\sqrt{r})(v+\sqrt{r+\gamma})}e^{tv^2-v|z|} \\
 &~~ + \int_0^{\min(\sqrt{r-\beta},\frac{|z|}{2t})} dv~ \frac{v^2\sqrt{r-\beta-v^2}}{(v-a)(v+\sqrt{r})(v+\sqrt{r+\gamma})}e^{tv^2-v|z|}.
 \end{split}
 \label{mbbJ_a^A>}
\end{align}
It is clear from the above equation and Eq.~\eqref{J_a^A-F} that the contribution of the form $\mathbb{J}_{a, \zeta}^{\mathcal{A}^{(3)}}\left(x,t\right)$ from terms like $ \mathcal{J}_{a}^{\mathcal{A}}(x,t) $ in the third line of Eq.~\eqref{r-g-beta} gets partly cancelled with the fourth line of Eq.~\eqref{r-g-beta} and the remaining contribution in the form $\mathbb{J}_{a, \zeta}^{\mathcal{A}^{(3)}}\left(x,t\right)$ is given by:
\begin{align}
\text{Combined~}&\text{contribution~of~the~form~} \mathbb{J}_{a, \zeta}^{\mathcal{A}^{(3)}}\left(x,t\right)~\text{from~the~$3^{rd}$~and~$4^{th}$~line~in~Eq.~\eqref{r-g-beta}} \nonumber \\ 
&=  \int_{\min(\sqrt{r-\beta},\frac{|z|}{2t})}^{\sqrt{r-\beta}} dv~ \frac{v^2\sqrt{r-\beta-v^2}}{(v-a)(v+\sqrt{r})(v+\sqrt{r+\gamma})}e^{tv^2-v|z|}. 
\label{mbbJ_a^A>-f}
\end{align}
Note that the above integral has to be performed on a real line segment and is non zero only for $|z| < 2t\sqrt{r-\beta}$. This integral can again be computed using saddle point method as discussed in previous sections because $\beta + \gamma>0$ for $r> \beta$ (recall $r+\gamma=r^2/(4 \beta)$ from Eq. (\ref{z(x)})).

\end{document}